\documentclass[aps,prd,twocolumn,superscriptaddress,nofootinbib]{revtex4-2}
\usepackage{amsmath}
\usepackage{amssymb}
\usepackage{graphicx}

\newcommand{\nShiftCoeff}{0.362381}  % replay/r6/prep/closed/quadrature_pslq_mpmath_output.txt:36, R5 source, same digits: anc_staging/5_closed_forms/residues_branch_point_output.txt:8; coefficient -(3/2)<w>_P of n_f alpha_s^2/N_c in alpha_O - 1, Eq. (main) (derived); unrounded 0.3623812194219710465837347994156153373609; 6 significant digits, half up; sign - typed in the text
\newcommand{\nShiftCoeffShort}{0.362}  % replay/r6/prep/closed/quadrature_pslq_mpmath_output.txt:36, R5 source, same digits: anc_staging/5_closed_forms/residues_branch_point_output.txt:8; coefficient -(3/2)<w>_P of n_f alpha_s^2/N_c in alpha_O - 1, Eq. (main), short form (derived); unrounded 0.3623812194219710465837347994156153373609; 3 significant digits, half up; sign - typed in the text
\newcommand{\nShiftCoeffAbar}{3.57656}  % replay/r6/prep/closed/quadrature_pslq_mpmath_output.txt:37, R5 source, same digits: anc_staging/5_closed_forms/residues_branch_point_output.txt:8; coefficient -(3 pi^2/2)<w>_P of n_f abar_s^2/N_c^3 in alpha_O - 1, Eq. (main) (derived); unrounded 3.57655927807921400271651094901972506272; 6 significant digits, half up; sign - typed in the text
\newcommand{\nWAvg}{0.2415874796}  % replay/r6/prep/closed/quadrature_pslq_mpmath_output.txt:35, R5 source, same digits: anc_staging/5_closed_forms/residues_branch_point_output.txt:7; pair average <w>_P of the function w of Eq. (w) from its closed form, Eq. (wP) (derived); unrounded 0.2415874796146473643891565329437435582406; 10 significant digits, half up
\newcommand{\nWAvgShort}{0.242}  % replay/r6/prep/closed/quadrature_pslq_mpmath_output.txt:35, R5 source, same digits: anc_staging/5_closed_forms/residues_branch_point_output.txt:7; pair average <w>_P, short form (derived); unrounded 0.2415874796146473643891565329437435582406; 3 significant digits, half up

\newcommand{\nAlphaRef}{0.2}  % anc_staging/7_nlo_eigenvalue_fig2/nlo_scale_dependence_output.txt:1, replay/r6/fig3_r6/fig2_nlo_n1.py:29; fixed coupling alpha_s(q^2) of the numerical examples (N_c = n_f = 3) (input); unrounded 0.2; as printed by the program
\newcommand{\nShiftCentral}{0.0145}  % anc_staging/7_nlo_eigenvalue_fig2/nlo_scale_dependence_output.txt:1; shift alpha_O - 1 at alpha_s(q^2) = 0.2, N_c = n_f = 3, mu^2 = q^2 (computed); unrounded -0.0144952487769; 3 significant digits, half up; sign - typed in the text
\newcommand{\nShiftMuHigh}{0.0101}  % anc_staging/7_nlo_eigenvalue_fig2/nlo_scale_dependence_output.txt:5; shift at nu = 0, and branch point, for mu^2 = 4q^2 at fixed alpha_s(q^2) = 0.2 (computed); unrounded -0.0100901445652; 3 significant digits, half up; sign - typed in the text
\newcommand{\nShiftMuLowOrigin}{0.0226}  % anc_staging/7_nlo_eigenvalue_fig2/nlo_scale_dependence_output.txt:2; eigenvalue at nu = 0 for mu^2 = q^2/4 at fixed alpha_s(q^2) = 0.2 (computed); unrounded -0.0225681881435; 3 significant digits, half up; sign - typed in the text
\newcommand{\nShiftMuLowMax}{0.0225}  % anc_staging/7_nlo_eigenvalue_fig2/nlo_scale_dependence_output.txt:2; branch point for mu^2 = q^2/4: the maximum of the eigenvalue, at nu = 0.10518489 (computed); unrounded -0.0224681615305; 3 significant digits, half up; sign - typed in the text
\newcommand{\nNuMaxLow}{0.1}  % anc_staging/7_nlo_eigenvalue_fig2/nlo_scale_dependence_output.txt:2; position nu of the maximum for mu^2 = q^2/4 ("near") (computed); unrounded 0.10518489; 1 significant digit
\newcommand{\nExcessMaxLow}{10^{-4}}  % anc_staging/7_nlo_eigenvalue_fig2/nlo_scale_dependence_output.txt:2; excess of the eigenvalue at its maximum over its value at nu = 0 for mu^2 = q^2/4 ("about") (computed); unrounded 0.000100027; nearest power of ten
\newcommand{\nMuThreshold}{0.27}  % anc_staging/7_nlo_eigenvalue_fig2/nlo_scale_dependence_output.txt:31; scale mu^2/q^2 above which the maximum stays at nu = 0, at fixed alpha_s(q^2) = 0.2 ("about") (computed); unrounded 0.27336409; 2 significant digits
\newcommand{\nCurvCancelPercent}{63}  % anc_staging/7_nlo_eigenvalue_fig2/nlo_eigenvalue_n1_output.txt:87; percentage of the LO nu^2 coefficient at nu = 0 that the NLO terms cancel at alpha_s = 0.2, mu^2 = q^2 (computed); unrounded -0.628450994619; percent, nearest integer

\newcommand{\nAreaZero}{0.55}  % anc_staging/8_pair_distribution_fig1/fig1_pairs_checks.txt:21; area under the n' = 0 curve of Fig. 2 (share of the n' = 0 waves) (computed); unrounded 0.55460491505380497912; 2 decimal places, half up
\newcommand{\nAreaTwo}{0.44}  % anc_staging/8_pair_distribution_fig1/fig1_pairs_checks.txt:22; area under the n' = +-2 curve of Fig. 2 (computed); unrounded 0.43649460907012428912; 2 decimal places, half up
\newcommand{\nHighSpinPercentWord}{one}  % anc_staging/8_pair_distribution_fig1/fig1_pairs_checks.txt:26; bound in words on the percentage of the Born normalization in the waves with even |n'| >= 4 ("less than one percent") (computed); unrounded 0.00890047587607073; bound: percentage rounded up to an integer, written as a word
\newcommand{\nWPomeron}{1.080}  % anc_staging/8_pair_distribution_fig1/fig1_pairs_checks.txt:44; w(0,1/2) = 11 pi/32, the value of the function w of Eq. (w) that gives the non-planar quark term of the NLO pomeron intercept (computed); unrounded 1.07992247467149; 4 significant digits, half up

\newcommand{\nNuEdge}{3}  % replay/r6/fig3_r6/fig2_nlo_n1.py:107; right end nu of the lower panel of Fig. 3, the plotted range of the non-planar quark contribution (and of the gaps followed in task E6) (input); unrounded 3.0; as set in the program
\newcommand{\nCurveDigits}{40}  % replay/r6/fig3_r6/fig2_nlo_n1.py:27; working precision in digits of the curves of Fig. 3 (input); unrounded 40; as set in the program
\newcommand{\nCurvQuark}{0.608}  % replay/r7e/package/anc/09_closed_forms/fnu_expansion_output.txt:6, same file as the R5/R6 source anc_staging/5_closed_forms/fnu_expansion_output.txt:6; the quark nu^2 coefficient, the coefficient of nu^2 in -(3/2)<w>_P(nu) (Eq. (F4)), in units of n_f alpha_s^2/N_c: in R7f the value of the new symbol for this coefficient, used in Eqs. (omegaexp) and (Fexp) and given once with ``approximately'' (R7f block, entry EQ46; audit item 9) (computed); unrounded 0.6082093293863053679157593742185107986753960040179112339689; closed form in zeta values (11 terms) in line 7 of the same file, kept in the ancillary files; 3 significant digits, half up

\newcommand{\nLevelQthreeA}{0.33155}  % replay/r6/prep/levels/levels_nu0_inner_output.txt:74, R5 source, same digits: anc_staging/6_three_gluon_levels/levels_monodromy_output.txt:6; Table I: q_3 (real) of the level with quasimomentum 1 x 2 pi/3 (computed); unrounded 0.33155402856950904786; 5 decimal places, half up
\newcommand{\nLevelOmegaA}{0.38427}  % replay/r6/prep/levels/levels_nu0_inner_output.txt:74, R5 source, same digits: anc_staging/6_three_gluon_levels/levels_monodromy_output.txt:6; Table I: omega/abar_s = -E_3/4 of that level (computed); unrounded E_3 = 1.5370810031239667591, E_3/4 = 0.38427025078099168977; 5 decimal places, half up; sign - typed in the text
\newcommand{\nLevelQthreeReB}{0.47285}  % replay/r6/prep/levels/levels_nu0_output.txt:77, replay/r6/prep/levels/levels_nu0_output.txt:78, R5 source, same digits: anc_staging/6_three_gluon_levels/levels_monodromy_output.txt:7; Table I: real part of q_3 of the complex level (computed); unrounded 0.47285041882780614265; 5 decimal places, half up
\newcommand{\nLevelQthreeImB}{0.72552}  % replay/r6/prep/levels/levels_nu0_output.txt:77, replay/r6/prep/levels/levels_nu0_output.txt:78, R5 source, same digits: anc_staging/6_three_gluon_levels/levels_monodromy_output.txt:7; Table I: imaginary part of q_3 of the complex level (computed); unrounded 0.72551508011928004664; 5 decimal places, half up
\newcommand{\nLevelOmegaB}{1.60466}  % replay/r6/prep/levels/levels_nu0_output.txt:77, replay/r6/prep/levels/levels_nu0_output.txt:78, R5 source, same digits: anc_staging/6_three_gluon_levels/levels_monodromy_output.txt:7; Table I: omega/abar_s = -E_3/4 of that level (computed); unrounded E_3 = 6.4186366393495307702, E_3/4 = 1.6046591598373826925; 5 decimal places, half up; sign - typed in the text
\newcommand{\nLevelQthreeC}{1.36802}  % replay/r6/prep/levels/levels_nu0_output.txt:79, R5 source, same digits: anc_staging/6_three_gluon_levels/levels_monodromy_output.txt:5; Table I: q_3 (real) of the level with quasimomentum 0 x 2 pi/3 (computed); unrounded 1.3680197663563365492; 5 decimal places, half up
\newcommand{\nLevelOmegaC}{2.02888}  % replay/r6/prep/levels/levels_nu0_output.txt:79, R5 source, same digits: anc_staging/6_three_gluon_levels/levels_monodromy_output.txt:5; Table I: omega/abar_s = -E_3/4 of that level (computed); unrounded E_3 = 8.1155173330199150531, E_3/4 = 2.0288793332549787633; 5 decimal places, half up; sign - typed in the text
\newcommand{\nLevelQthreeD}{2.20681}  % replay/r6/prep/levels/levels_nu0_output.txt:80, R5 source, same digits: anc_staging/6_three_gluon_levels/levels_monodromy_output.txt:8; Table I: imaginary q_3 of the level with quasimomentum 0 x 2 pi/3 (computed); unrounded 2.2068148059766997409; 5 decimal places, half up
\newcommand{\nLevelOmegaD}{2.52885}  % replay/r6/prep/levels/levels_nu0_output.txt:80, R5 source, same digits: anc_staging/6_three_gluon_levels/levels_monodromy_output.txt:8; Table I: omega/abar_s = -E_3/4 of that level (computed); unrounded E_3 = 10.115394317133955228, E_3/4 = 2.5288485792834888071; 5 decimal places, half up; sign - typed in the text
\newcommand{\nGapBose}{2.03}  % replay/r6/prep/levels/levels_nu0_output.txt:79, R5 source, same digits: anc_staging/6_three_gluon_levels/levels_monodromy_output.txt:5; gap omega/abar_s at nu = 0 between the BLV eigenvalue and the nearest level of zero quasimomentum (the symmetry of the BLV state) (computed); unrounded 2.0288793332549787633; 3 significant digits, half up
\newcommand{\nQthreeCut}{3}  % replay/r5/numbers_r5.tex:27; bound on |q_3| of the levels listed in Table I (choice); unrounded 3; choice of the paper, carried from R5
\newcommand{\nOmegaCut}{2.6}  % replay/r5/numbers_r5.tex:28; lower bound on omega/abar_s of the levels listed in Table I (choice); unrounded 2.6; choice of the paper, carried from R5; it lies between row D and row E
\newcommand{\nQthreeCutWide}{6}  % replay/r6/prep/levels/NOTE.md:13, replay/r6/prep/levels/levels_nu0_output.txt:1; bound on |q_3| of the widened search for levels at nu = 0 (task E6) (input); unrounded 6; as stated by the preparation; the grid of the program covers it

\newcommand{\nGapNuOne}{0.959}  % replay/r6/prep/levels/analyze_all_output.txt:6, replay/r6/prep/levels/scan_nu_1.0_output.txt:7; smallest gap omega/abar_s from the BLV eigenvalue to a level of zero quasimomentum over 0 <= nu <= 1 (upper panel of Fig. 3), at nu = 1 (computed); unrounded 0.9594436963115891; 3 significant digits, half up
\newcommand{\nGapNuThree}{0.0402}  % replay/r6/prep/levels/check_nu3_monodromy_output.txt:1, replay/r6/prep/levels/analyze_all_output.txt:6; smallest gap omega/abar_s from the BLV eigenvalue to a level of zero quasimomentum over 0 <= nu <= 3 (lower panel of Fig. 3), at nu = 3 (computed); unrounded 0.04021791115700481; 3 significant digits, half up
\newcommand{\nNuCrossSame}{3.35}  % replay/r6/prep/levels/beyond_nu3_output.txt:19, replay/r6/prep/levels/check_nu3_monodromy_output.txt:2; nu at which the nearest level of zero quasimomentum (row C) crosses the BLV eigenvalue, just beyond the plotted range (computed); unrounded 3.34837; 3 significant digits, half up
\newcommand{\nNuCrossOtherA}{0.64}  % replay/r6/prep/levels/analyze_all_output.txt:3; nu at which the level of quasimomentum +-2 pi/3 from row A crosses the BLV eigenvalue (a level the quark term cannot mix with the BLV state) (computed); unrounded 0.6388085532256681; 2 significant digits, half up
\newcommand{\nNuCrossOtherB}{1.80}  % replay/r6/prep/levels/analyze_all_output.txt:4; nu at which the trajectory of quasimomentum +-2 pi/3 through row B crosses the BLV eigenvalue (computed); unrounded 1.8034683062871026; 3 significant digits, half up
\newcommand{\nShiftGapPercent}{4}  % anc_staging/7_nlo_eigenvalue_fig2/nlo_scale_dependence_output.txt:1, replay/r6/prep/levels/levels_nu0_output.txt:79; shift at nu = 0 as a percentage of the gap nGapBose, at alpha_s = 0.2, N_c = n_f = 3 ("about") (derived); unrounded 0.0374083090462; percent, nearest integer; abar_s = alpha_s N_c/pi

\newcommand{\nLevelDiffPercentA}{0.02}  % sources_arxiv/src/hep-th_0204124/DKKM-revised.tex:2893, replay/r6/prep/levels/levels_nu0_inner_output.txt:74; relative difference, in percent, of the energy printed in Table 4 of hep-th/0204124 from E_3 of row A (derived); unrounded 0.0182816; 1 significant digit, half up
\newcommand{\nLevelDiffPercentC}{0.09}  % sources_arxiv/src/hep-th_0204124/DKKM-revised.tex:2927, replay/r6/prep/levels/levels_nu0_output.txt:79; relative difference, in percent, of the energy printed in Table 5 of hep-th/0204124 from E_3 of row C (derived); unrounded 0.0926291; 1 significant digit, half up
\newcommand{\nOddStateCoeff}{0.247}  % sources_arxiv/src/hep-th_9802100/oso2.tex:381; coefficient c in the intercept 1 - c alpha_s N_c/pi of the LO C-odd state of Janik and Wosiek (literature); unrounded -0.24717; 3 significant digits, half up; sign - typed in the text

\newcommand{\nKqDigits}{20}  % replay/r7e/package/anc/07_quark_kernel_eigenvalue/kq_eigenvalue.py:27 (mp.mp.dps = 20); working precision in digits of the arithmetic of the integrand of Eq. (Blambda); the Gauss-Legendre nodes and weights come from numpy.polynomial.legendre.leggauss in double precision (gl(), replay/r7e/package/anc/07_quark_kernel_eigenvalue/kq_eigenvalue.py:106-107), which limits the rule to about 1e-15 relative (audit item 11, A7.2b, dc6 C10) (input); unrounded 20; as set in the program
\newcommand{\nKqNuMin}{0.001}  % anc_staging/3_quark_kernel_eigenvalue/kq_eigenvalue_output.txt:18; smallest pair label nu' at which Eq. (Blambda) was evaluated (input); unrounded 0.001; as printed by the program
\newcommand{\nKqNuMax}{1.1}  % anc_staging/3_quark_kernel_eigenvalue/kq_eigenvalue_output.txt:8; largest pair label nu' at which Eq. (Blambda) was evaluated (input); unrounded 1.1; as printed by the program
\newcommand{\nRatioNuA}{0.3}  % anc_staging/4_ratio_quadrature/disk_grid_ratio_48_output.txt:2; first label nu at which the ratio quadrature was repeated for Eq. (ratio) (input); unrounded 0.3; as printed by the program
\newcommand{\nRatioNuB}{0.8}  % anc_staging/4_ratio_quadrature/disk_grid_ratio_48_output.txt:3; second label nu of the repeated ratio quadrature (input); unrounded 0.8; as printed by the program
\newcommand{\nFnuNuA}{0.3}  % anc_staging/5_closed_forms/fnu_quadrature_check_output.txt:1; label nu at which Eq. (Fnu) was compared with direct quadrature of <w>_P(nu) (input); unrounded 0.3; as printed by the program
\newcommand{\nFnuNuB}{0.8}  % anc_staging/5_closed_forms/fnu_quadrature_check_output.txt:2; label nu at which Eq. (Fnu) was compared with direct quadrature of <w>_P(nu) (input); unrounded 0.8; as printed by the program
\newcommand{\nFnuNuC}{1.7}  % anc_staging/5_closed_forms/fnu_quadrature_check_output.txt:3; label nu at which Eq. (Fnu) was compared with direct quadrature of <w>_P(nu) (input); unrounded 1.7; as printed by the program
\newcommand{\nFnuNuD}{3.0}  % anc_staging/5_closed_forms/fnu_quadrature_check_output.txt:4; label nu at which Eq. (Fnu) was compared with direct quadrature of <w>_P(nu) (input); unrounded 3.0; as printed by the program
\newcommand{\nLevelDigits}{40}  % replay/r6/prep/levels/levels_nu0.py:28, replay/r6/prep/levels/levels_nu0.py:29; working precision in digits of the levels at h = 1 (task E6: roots refined and carried to h = 1) (input); unrounded 40; as set in the program
\newcommand{\nLevelDigitsHigh}{50}  % replay/r6/prep/levels/levels_nu0.py:30; raised working precision in digits at which the levels at h = 1 were checked (task E6; roots and energies stable in every printed digit) (input); unrounded 50; as set in the program
\newcommand{\nFollowStep}{0.005}  % replay/r6/prep/levels/follow_np.py:62; step in nu with which every level found at nu = 0 was followed to nu = 3 (task E6, double precision) (input); unrounded 0.005; as set in the program
\newcommand{\nFollowDigits}{30}  % replay/r6/prep/levels/scan_nu.py:20; working precision in digits at which the levels found at fixed nu were confirmed (task E6) (input); unrounded 30; as set in the program
\newcommand{\nLeftEigNu}{0.3}  % replay/r6/prep/transfer/t4_output_h0.25.txt:8; nonzero label nu at which eq:lefteig was checked at n = 1; unrounded 0.3; as printed
\newcommand{\nAgreeLeftEig}{3}  % replay/r6/prep/transfer/t4_output_h0.25.txt:10 (in-span control), :11 (off-span); significant digits to which eq:lefteig held for T(D_xy,D_zz',D_zz') at n = 1, nu = \nLeftEigNu, set by the in-span control of the same run; unrounded off-span 2.30e-05, in-span control 2.93e-04, relative to 2 pi chi(1,nu); floor(-log10 r), r the larger of the two
\newcommand{\nQthreeLargeMax}{44}  % replay/r7c/inputs/levels_large/large_q3_levels_output.txt:11; largest |q3| among those levels, rounded down; unrounded 44.069519; floor
\newcommand{\nOmegaLargeTop}{3.82}  % replay/r7c/inputs/levels_large/large_q3_levels_output.txt:4; minus the highest LO eigenvalue omega/abar_s among those levels of zero quasimomentum (at |q3| = 8.1263), rounded down, so that every one of them lies at least this far below the BLV eigenvalue zero; unrounded -3.8271863971; magnitude rounded down to two decimals (a lower bound)
\newcommand{\nDevMaxOne}{0.07}  % replay/r7c/inputs/levels_large/large_q3_levels_output.txt:17; upper bound on |E3 - 4 ln|q3| - 12 gamma_E| over all levels found at (1,0) with |q3| > 1 (15 levels, R6 and R7c, every quasimomentum); unrounded 0.064528 (largest, at |q3| = 1.3680); rounded up to two decimals
\newcommand{\nFitQMin}{1}  % gen_numbers_r7c.py (THR1); lower threshold on |q3| of the deviation bound nDevMaxOne; unrounded 1; parameter of the program
\newcommand{\nDevMaxSix}{0.003}  % replay/r7c/inputs/levels_large/large_q3_levels_output.txt:23; upper bound on |E3 - 4 ln|q3| - 12 gamma_E| over the levels found with |q3| > 6 (9 levels); unrounded 0.002216 (largest, at |q3| = 7.2049); rounded up to three decimals

\begin{document}

\title{Non-planar quark loops lower the odderon branch point of Bartels, Lipatov and Vacca at next-to-leading order and fixed coupling}

\author{Alex Prygarin}
\email[Corresponding author, ]{alexanderp@ariel.ac.il}
\affiliation{Department of Physics, Ariel University, Ariel 40700, Israel}
\author{Claudelle Capasia Madjuogang Sandeu}
\email{claudell.madjuoga@msmail.ariel.ac.il}
\affiliation{Department of Physics, Ariel University, Ariel 40700, Israel}
\author{Karam Shekh Yusuf}
\email{yusufk@ariel.ac.il}
\affiliation{Department of Physics, Ariel University, Ariel 40700, Israel}

\begin{abstract}
We calculate the first-order shift of the branch point of the three-gluon odderon cut of Bartels, Lipatov and Vacca from the next-to-leading-order dipole equation at weak fixed coupling.
The calculation uses dilute three-gluon exchange without saturation in QCD with at least three colors and few enough massless fundamental quark flavors for asymptotic freedom.
We assume the three-gluon state through which the cut couples to a dipole at its branch point shares its leading-order eigenvalue with no other three-gluon state matching its conformal labels and cyclic permutation symmetry.
Non-planar quark loops lower the branch point below one in proportion to the number of flavors by attaching one gluon to the parent dipole and two to the dipole of the emitted quark-antiquark pair.
The state these loops produce is not a superposition of the three-gluon states of single color dipoles and gives the whole shift of the branch point, which at this order stays at one for gluons and adjoint matter that keeps the theory asymptotically free.
The lowering by fundamental quarks makes amplitudes governed by the cut fall faster with energy.
\end{abstract}

\maketitle

\section{Introduction}
\label{sec:intro}
D0 and TOTEM~\cite{D0TOTEM:2021} found evidence for exchange odd under charge conjugation ($C$-odd) by comparing elastic $pp$ and $p\bar p$ scattering. The collaborations reported observing $C$-odd exchange after including TOTEM's total cross section and the ratio of the real to the imaginary part of the forward amplitude. These measurements determine neither the odderon intercept nor the exchanged state. At weak coupling, Kovchegov, Szymanowski and Wallon~\cite{KovchegovSzymanowskiWallon:2004} related leading-order (LO) odderon exchange between a dipole and a dilute target to the three-gluon cut of Bartels, Lipatov and Vacca (BLV)~\cite{BartelsLipatovVacca:2000}. The LO eigenvalue of the BLV state at conformal spin one has a maximum of zero, so the cut starts at angular momentum one~\cite{BartelsLipatovVacca:2000}. The contribution of the cut to the $C$-odd dipole amplitude then has no exponential fall in rapidity. A branch point below one gives exponential decay and a faster fall with energy. The $C$-odd dipole amplitude enters single transverse-spin asymmetries through the mechanism of Ref.~\cite{KovchegovSievert:2012}.

It remains open whether the BLV branch point stays at angular momentum $j=1$ at next-to-leading order (NLO) in QCD with fundamental quarks beyond the planar limit. The conformal spin $n$ and real $\nu$ label the conformal eigenfunctions and their LO eigenvalues, and the LO maximum for the BLV state at $n=1$ lies at $\nu=0$. Caron-Huot~\cite{CaronHuot:2015} obtained an intercept of one to all orders in the planar limit and asked whether fundamental matter loops move the branch point from $j=1$ at NLO. Sta\'sto~\cite{Stasto:2009} argued from the $n=1$ singlet BFKL eigenvalue for an intercept of one to all orders and called for an explicit NLO evaluation of the odderon. The linearized form of the NLO dipole equation~\eqref{eq:nloterms}, used in Ref.~\cite{Kovchegov:2013}, keeps the quark loop only in its two-gluon, $C$-even form and gives an intercept of one up to corrections of order $\alpha_s^3$. Bartels and Vacca~\cite{BartelsVacca:2013} assumed a generalized NLO bootstrap and obtained an intercept of one from the same singlet eigenvalue in $\mathcal N=4$ super Yang-Mills theory and planar QCD. They wrote, \lq\lq without further direct QCD calculations we cannot conclude that also in QCD, for finite $N_c$, the NLO Odderon has intercept one''~\cite{BartelsVacca:2013}. In SU(3) the quark part of the $C$-odd equation contains non-dipole operators at three-gluon order, three-quark Wilson loops at three distinct points~\cite{BalitskyGrabovsky:2015}.

In this paper we calculate the shift of the BLV branch point due to non-planar fundamental quark loops to first order in the NLO dipole kernel~\cite{BalitskyChirilli:2008}.
The theory is QCD with $N_c\geq3$ colors and $n_f$ massless fundamental flavors with $n_f<11N_c/2$.
At zero momentum transfer we keep cubic order in the target field without saturation. The coupling stays weak and fixed during evolution, and the one-loop beta function terms are evaluated at one renormalization scale.

We write $A_x$ for the target field at the transverse point $x$. The field difference between the ends $x$ and $y$ of a dipole is $D_{xy}=A_x-A_y$. We contract its color components $D_{xy}^a$ with the symmetric color tensor $d^{abc}$ to obtain the cubic imaginary part of the unnormalized dipole trace. The resulting dipole form~\cite{HattaIancuItakuraMcLerran:2005}
\begin{equation}
C(D_{xy})=-\frac{1}{24}\,d^{abc}D_{xy}^aD_{xy}^bD_{xy}^c
\label{eq:Cdipole}
\end{equation}
describes three reggeized gluons in the symmetric color state $d^{abc}$, each at one end of the dipole. Let $W$ be the cubic form of the projectile's three-gluon state. The local term $C(A_x)$ is the cubic form of Eq.~\eqref{eq:Cdipole} evaluated on the field at $x$ alone, so all three gluons are at that point. Its signed coefficient in $W$ is the local coefficient $W_x$. A single dipole has $W_x=1$ and $W_y=-1$. A test function $g$ of transverse position defines the functional $L_g$,
\begin{equation}
L_g[W]=\sum_x g(x)\,W_x.
\label{eq:Lgdef}
\end{equation}

LO emission sends a dipole to a superposition of dipoles. It conserves $L_g$ for every test function because the local coefficients cancel at the daughters' shared end and the virtual term subtracts the parent. For the choice of $g$ in Sec.~\ref{sec:formalism}, $L_g$ of a dipole is proportional to its overlap with the BLV state at the branch point. The constant is the same for every dipole, so the relation holds throughout a superposition. Adjoint matter and every NLO term except the non-planar quark loop also send a dipole to a superposition of dipoles and conserve $L_g$,
so they leave the overlap at the branch point unchanged in rapidity. The unchanged overlap gives no first-order correction to the eigenvalue there. The branch point therefore stays at one for gluons and for adjoint matter that keeps the theory asymptotically free.

An emitted gluon can split into a quark-antiquark pair. For the non-planar loop the Fierz identity gives quadrupoles, single traces of four Wilson lines with a factor of $1/N_c$. These quadrupoles produce a three-gluon state in which one exchanged gluon stays on the parent dipole and two attach to the dipole of the $q\bar q$ pair. The three gluons cannot all meet at one point when the four dipole ends are distinct, so $L_g$ vanishes for this state. Its overlap with the BLV state is nevertheless positive, so the state is not a superposition of the three-gluon states of single color dipoles. The loop multiplies this positive overlap by its negative color coefficient and thus lowers the BLV eigenvalue at the branch point.

We derive the LO conservation law and the sign of the correction in Secs.~\ref{sec:formalism} and~\ref{sec:shift}, compute its size in Sec.~\ref{sec:result} and locate the cut in Sec.~\ref{sec:location}.

The maximum of the real part of the $n=1$ eigenvalue over real $\nu$ fixes the branch point $j=\alpha_O$. At weak coupling the maximum lies at $\nu=0$. We evaluate the contribution of a gluon pair in the BLV state using the non-planar quark term on a color-singlet pair in the NLO BFKL eigenvalue of Kotikov and Lipatov~\cite{KotikovLipatov:2000}. This term is of order $n_f/N_c^3$ in units of $\bar\alpha_s^2$, with $\bar\alpha_s=\alpha_s N_c/\pi$. Its dependence on the conformal labels of the gluon pair is given by $w$ in Eq.~\eqref{eq:w}. Averaging $w$ over those labels for one gluon pair in the BLV state at $(n,\nu)=(1,0)$ gives $\langle w\rangle_P$. Each of the three gluon pairs contributes twice the non-planar quark term on a color-singlet pair with this average, giving
\begin{align}
\alpha_O-1&=-\frac{3n_f\alpha_s^2}{2N_c}\,\langle w\rangle_P
=-\nShiftCoeff\,\frac{n_f\alpha_s^2}{N_c}\nonumber\\
&=-\nShiftCoeffAbar\,\frac{n_f}{N_c^3}\,\bar\alpha_s^2.
\label{eq:main}
\end{align}
We show that fundamental quarks lower $\alpha_O$ if no other three-gluon state at $(1,0)$ with the same cyclic permutation symmetry has the BLV eigenvalue. The non-planar quark loop produces a state that is not a superposition of dipoles, and this state accounts for the entire shift of the branch point.
For gluons the BLV branch point stays at one, a result that also holds for adjoint matter that keeps the theory asymptotically free.
The shift is $-\nShiftCentral$ at $\alpha_s=\nAlphaRef$ and $N_c=n_f=3$ for the scale and scheme of Sec.~\ref{sec:location}. The branch point may return to one if the coupling runs during evolution, although we still expect a scale-dependent suppression of the amplitude by the non-planar loop (Sec.~\ref{sec:concl}). The result answers the question of Caron-Huot and of Bartels and Vacca for fundamental quarks through NLO at fixed coupling if no other three-gluon state with the same labels and cyclic symmetry shares the BLV eigenvalue at the branch point.

\section{Three-gluon exchange and the conservation of $L_g$}
\label{sec:formalism}
We first show how to obtain the overlap at the LO branch point from the local coefficients of the dipole form in Eq.~\eqref{eq:Cdipole}. We use these coefficients to follow the overlap through an emission. Their conservation will let us determine which terms of the NLO equation leave the overlap unchanged.

The LO $C$-odd dipole amplitude obeys the BFKL equation at odd conformal spin for every $N_c\geq3$~\cite{HattaIancuItakuraMcLerran:2005}. At fixed $(n,\nu)$ the component of the projectile dipole's three-gluon state is proportional to the BLV state~\cite{KovchegovSzymanowskiWallon:2004}, a $C$-odd LO eigenstate of three reggeized gluons in the symmetric color state $d^{abc}$. The LO eigenvalue $\bar\alpha_s\chi(n,\nu)$ of the BLV state determines its evolution factor $\exp[\bar\alpha_s\chi(n,\nu)Y]$ over rapidity $Y$. The characteristic function involves the digamma function $\psi$,
\begin{equation}
\chi(n,\nu)=2\psi(1)-\psi\Big(\frac{1+|n|}{2}+i\nu\Big)-\psi\Big(\frac{1+|n|}{2}-i\nu\Big).
\label{eq:chi}
\end{equation}
We use complex transverse coordinates. The conformal eigenfunctions diagonalize the two Casimir operators of the conformal M\"obius group of the transverse plane. Their weights are $h=(1+n)/2+i\nu$ and $\bar h=(1-n)/2+i\nu$. The dependence on the dipole ends $x,y$ and the reference point $x_0$ is
\begin{align}
E^{n,\nu}(x,y;x_0)&=\Big[\frac{x-y}{(x-x_0)(y-x_0)}\Big]^{h}\nonumber\\
&\quad\times\Big[\frac{\bar x-\bar y}{(\bar x-\bar x_0)(\bar y-\bar x_0)}\Big]^{\bar h},
\label{eq:Enu}
\end{align}
where a bar on a coordinate denotes complex conjugation. The overlap $\Pi_{n,\nu}$ extracts the BLV component of a dipole-dipole amplitude $F$. We integrate over the target ends $u,v$ against the conformal eigenfunction itself,
\begin{equation}
\Pi_{n,\nu}[F](x,y;x_0)=\int\frac{d^2u\,d^2v}{|u-v|^4}\,E^{n,\nu}(u,v;x_0)F(x,y;u,v).
\label{eq:Pidef}
\end{equation}
For a conformally invariant amplitude the remaining dependence on the projectile ends is that of the same eigenfunction.
The reference point remains common to target and projectile after the integration over the target ends. The coefficient multiplying the projectile eigenfunction measures the part of the exchange that couples to the BLV state with those labels.

We describe the three gluons of the projectile by expanding its Wilson lines $U_x=\exp(iA_x)$ in the target field $A_x=A_x^at^a$, which includes the coupling. The $C$-odd part of an unnormalized trace is $i$ times its imaginary part.
The generators are traceless, and the trace of two generators is real. The first imaginary term thus comes from three field insertions. The imaginary part selects the symmetric color tensor $d^{abc}$ at this order, so the insertion points enter through the signed sum of their fields.
Assigning a sign $\sigma_i$ of $+1$ to a line and $-1$ to its Hermitian conjugate, we obtain
\begin{equation}
C(D)=-\frac{1}{24}\,d^{abc}D^aD^bD^c,\qquad D=\sum_i\sigma_i\,A_{x_i},
\label{eq:cubic}
\end{equation}
which reduces to Eq.~\eqref{eq:Cdipole} for a single dipole. The BLV state~\cite{BartelsLipatovVacca:2000} is a superposition of dipoles, a linear combination of the forms $C(D_{xy})$ for different pairs of ends. This expansion distinguishes the BLV state from non-dipole forms, cubic forms for which no such superposition exists~\cite{BalitskyGrabovsky:2015}.

The symmetric trilinear form $T$, normalized by $T(D,D,D)=C(D)$, also allows gluons on different dipoles. To find the overlap with the BLV state we apply $\Pi_{n,\nu}$ to the Born exchange of this form. Each argument $D_{x_i x_j}$ contributes a one-gluon exchange $l$ with the same target dipole.
After omission of a common color and coupling factor the exchange is
\begin{equation}
l(x,y;u,v)=\ln\frac{|x-v|\,|y-u|}{|x-u|\,|y-v|}.
\label{eq:l}
\end{equation}
The opposite signs of the fields at the two ends of each dipole enter this exchange at both the projectile and the target. Exchanging the ends of either dipole therefore reverses its sign. The product of the three exchanges retains the information about which dipole each gluon came from.
The non-planar quark loop produces cubic forms with gluons on different dipoles. To evaluate the functional $L_g$ of Eq.~\eqref{eq:Lgdef} on any cubic form, we extend it to the trilinear form.
A trilinear term contributes at a point only when all three arguments contain the field at that point.
The overlap with the BLV state uses the full Born exchange, including contributions from fields at several positions. The functional depends only on local terms of the projectile cubic form and can be evaluated from that form without computing its exchange with a target dipole.
Let $Q_{k,x}$ be the coefficient of $A_x$ in the argument $D_k$. The local coefficient of $T(D_1,D_2,D_3)$ is then $W_x=Q_{1,x}Q_{2,x}Q_{3,x}$, so
\begin{equation}
L_g\Big[T(D_1,D_2,D_3)\Big]=\sum_{x}g(x)\,Q_{1,x}\,Q_{2,x}\,Q_{3,x}.
\label{eq:Lg}
\end{equation}
For emission integrals we integrate the functional of the integrand with the symmetric prescription of Appendix~\ref{sm:charge}. The two daughters of an emission at $z$ have opposite local coefficients at their shared end. The virtual subtraction cancels the remaining contributions at the original ends, so
\begin{align}
&L_g\big[C(D_{xz})+C(D_{zy})-C(D_{xy})\big]\nonumber\\
&\quad=g(x)-g(z)+g(z)-g(y)-g(x)+g(y)=0.
\label{eq:telescope}
\end{align}
The same cancellation occurs at each shared end of the dipoles joined end to end in the planar argument of Caron-Huot~\cite{CaronHuot:2015}.
The cancellation holds for every test function because it takes place separately at each end. One choice of test function relates the conserved functional to the overlap with the BLV state at the branch point.
The choice for $L_g$ is $g(x)=-1/(x-x_0)$, for which the difference between the ends is the projectile conformal eigenfunction,
\begin{equation}
L_g\big[C(D_{xy})\big]=g(x)-g(y)=E^{1,0}(x,y;x_0).
\label{eq:LgDipole}
\end{equation}
Conformal covariance requires the same eigenfunction in the overlap with the BLV state. At $(1,0)$ we denote this overlap by $\Pi_{1,0}$. The Born exchange of that dipole is $l^3$, and the integral over the target in Appendix~\ref{sm:pairs} fixes the proportionality constant,
\begin{equation}
\Pi_{1,0}\big[l(x,y;u,v)^3\big]=6\pi^2\zeta(3)\,E^{1,0}(x,y;x_0).
\label{eq:Pi0dipole}
\end{equation}
The coefficient is independent of the dipole. Thus $\Pi_{1,0}$ is proportional to $L_g$ for any superposition of dipoles when $g$ is the test function of Eq.~\eqref{eq:LgDipole}.

We have expressed $L_g$ on a trilinear form through its local coefficients. The overlap with the BLV state at the branch point is proportional to this functional on superpositions of dipoles, with the test function of Eq.~\eqref{eq:LgDipole} and the constant in Eq.~\eqref{eq:Pi0dipole}. LO emission sends a dipole to a superposition of dipoles and conserves the functional by Eq.~\eqref{eq:telescope}, so the overlap stays constant in rapidity and the LO branch point stays at $j=1$. In the next section we compute the effect of one NLO insertion on this overlap.

\section{The first-order shift}
\label{sec:shift}
We now show that the non-planar quark loop is the only NLO term that can move the LO branch point at $j=1$. We follow the overlap that was constant under LO emission in Sec.~\ref{sec:formalism}. A conformally invariant NLO term can take the dipole into a non-dipole state in a single insertion. Its first-order contribution to the BLV eigenvalue is a ratio of overlaps with the BLV state, for which we need the left and right eigenvectors of the LO Hamiltonian. The component of the projectile dipole at fixed conformal labels gives the right eigenvector. The overlap $\Pi_{n,\nu}$ must act on the state after the insertion, so we need it to be a left eigenvector on non-dipole forms too.

To establish this property we let the LO kernel act on the target. Real emission at $z$ and the virtual subtraction act on a function $F$ of two dipole ends through
\begin{align}
(K_0F)(x,y)&=\int d^2z\,\frac{|x-y|^2}{|x-z|^2|z-y|^2}\nonumber\\
&\quad\times\big[F(x,z)+F(z,y)-F(x,y)\big].
\label{eq:K0}
\end{align}

Each gluon corresponds to one argument of the trilinear form $T$. The pair kernel $H_{ij}$ acts on the arguments for gluons $i,j$ while the third argument, the spectator, is held fixed. We normalize the kernel to the LO color-singlet pair kernel, so it reduces to $\bar\alpha_s K_0/(2\pi)$ acting on one dipole when both gluons belong to that dipole. The symmetric color tensor $d^{abc}$ puts each pair in the symmetric adjoint representation, the symmetric octet $8_s$ for SU(3). The LO color factor of a symmetric adjoint pair is half the color-singlet value, so the three-gluon Hamiltonian is
\begin{equation}
H_0=\tfrac12\big(H_{12}+H_{13}+H_{23}\big),
\label{eq:H0}
\end{equation}
with one term for each choice of spectator.
A gluon pair can be on one dipole even when the spectator belongs to another. In that case the two arguments on which the pair kernel acts already have the dipole form needed for the real emission and its subtraction. For a pair on different dipoles we need the same action on a product of two different field differences.

We polarize the quadratic form in the two fields, holding the spectator argument $D$ fixed. Every term on the right of the polarization identity has the two gluons on one dipole. Its coefficient is the same for the trilinear form and for the Born exchange,
{\footnotesize
\begin{align}
T(D_{xy},D_{zz'},D)&=-\tfrac12\big[T(D_{xz},D_{xz},D)\nonumber\\
&\quad-T(D_{xz'},D_{xz'},D)-T(D_{yz},D_{yz},D)\nonumber\\
&\quad+T(D_{yz'},D_{yz'},D)\big],\nonumber\\
l(x,y;u,v)\,l(z,z';u,v)&=-\tfrac12\big[l(x,z;u,v)^2\nonumber\\
&\quad-l(x,z';u,v)^2-l(y,z;u,v)^2\nonumber\\
&\quad+l(y,z';u,v)^2\big].
\label{eq:mixedpair}
\end{align}
}

Each pair on the right exchanges $l^2$ with the target. The logarithm in Eq.~\eqref{eq:l} is additive across a dipole splitting, so the real terms and the virtual subtraction combine into a product of the two daughter exchanges. The integrals converge absolutely without a regulator, and M\"obius invariance makes the result a function of the cross ratio. Interchanging projectile and target preserves both $l$ and the cross ratio. The kernel thus gives the same function acting on the projectile ends $(x,y)$ or the target ends $(u,v)$. Writing $p$ for the emission point on the target, we obtain
\begin{align}
K_0^{(x,y)}\,l(x,y;u,v)^2&=-2\int d^2z\,\frac{|x-y|^2}{|x-z|^2|z-y|^2}\nonumber\\
&\quad\times l(x,z;u,v)\,l(z,y;u,v)\nonumber\\
&=-2\int d^2p\,\frac{|u-v|^2}{|u-p|^2|p-v|^2}\nonumber\\
&\quad\times l(x,y;u,p)\,l(x,y;p,v)\nonumber\\
&=K_0^{(u,v)}\,l(x,y;u,v)^2.
\label{eq:transfer}
\end{align}
We denote the Born exchange between argument $m$ of the trilinear form and the target dipole by $l_m(u,v)$. A pair on different dipoles exchanges the product $l_i l_j$. Equation~\eqref{eq:mixedpair} resolves this product into squared exchanges. We can apply Eq.~\eqref{eq:transfer} to each one and then combine them with the same coefficients on the target side.

In the sum over pairs, $i<j$ labels the pair and $k$ the spectator. Emission at $p$ splits each logarithm into the sum of its exchanges with the daughters $(u,p)$ and $(p,v)$. The kernel now acts on the target ends of $l_i l_j$.
The integrand at $p$ for the pair kernel $H_{12}$ contains two real terms. They put both exchanges $l_1$ and $l_2$ on one daughter or both on the other. The virtual term subtracts their product on the parent. The spectator exchange $l_3(u,v)$ keeps both parent ends in all three terms. We omit the factor multiplying $K_0$ in the single-dipole kernel $\bar\alpha_s K_0/(2\pi)$ and the emission measure of Eq.~\eqref{eq:K0}, as in Eq.~\eqref{eq:pairsum}. The contribution of this pair includes the factor of one half from Eq.~\eqref{eq:H0} and is
\begin{align}
&\tfrac12\big[l_1(u,p)\,l_2(u,p)+l_1(p,v)\,l_2(p,v)\nonumber\\
&\qquad-l_1(u,v)\,l_2(u,v)\big]\,l_3(u,v).\nonumber
\end{align}
For a general pair the spectator exchange $l_k(u,v)$ multiplies the same difference between real and virtual terms. Summing over the three choices gives
\begin{align}
&\tfrac12\sum_{i<j}\big[l_i(u,p)\,l_j(u,p)+l_i(p,v)\,l_j(p,v)\nonumber\\
&\qquad-l_i(u,v)\,l_j(u,v)\big]\,l_k(u,v)\nonumber\\
&\quad=\prod_{m=1}^{3}l_m(u,p)+\prod_{m=1}^{3}l_m(p,v)-\prod_{m=1}^{3}l_m(u,v),
\label{eq:pairsum}
\end{align}
an identity that holds because each product involving both daughters occurs in two of the three pairs. This double occurrence cancels the factor of one half.
The real term for a single pair still has a gluon on the parent. Only after summing over the possible spectators do the two real terms have all three exchanges on the same daughter. The subtraction then acts on the exchange with the original target dipole.
The sum over pairs leaves $K_0$ acting on the target ends of the full Born exchange $F_W$ of the cubic form $W$. Thus $F_{H_0W}=(\bar\alpha_s/2\pi)K_0^{(u,v)}F_W$.

The conformal eigenfunction $E^{n,\nu}$ has eigenvalue $2\pi\chi(n,\nu)$ under $K_0$ (Appendix~\ref{sm:setting}). To use this eigenvalue in the overlap $\Pi_{n,\nu}[W]=\Pi_{n,\nu}[F_W]$, we must regulate the real terms and the virtual subtraction together because they diverge separately at the target ends. We exclude disks of radius $\eta$ around both ends and impose the same lower bound on the target dipole size.
The two distances from the emission point and the distance between the ends then obey the same restriction. The region is consequently symmetric in $u,v,p$. Exchanging the emission point with an end in a real term also preserves the product of the dipole measure and the emission density. The virtual subtraction needs no change of variables because the conformal eigenfunction and the Born exchange are already evaluated at the target ends.
These changes of variables put $K_0$ on the conformal eigenfunction. Using its eigenvalue and removing the common regulator as in Eq.~\eqref{eq:lefteig} of Appendix~\ref{sm:firstorder}, we obtain
\begin{equation}
\Pi_{n,\nu}\big[H_0W\big]=\bar\alpha_s\,\chi(n,\nu)\,\Pi_{n,\nu}[W],
\label{eq:lefteigid}
\end{equation}
so $\Pi_{n,\nu}$ remains a left eigenvector of $H_0$ on cubic forms built from dipole differences and their emission integrals, including non-dipole forms.

For the right eigenvector we resolve the projectile into BLV states, each with a singularity at $j=1+\omega$ for evolution eigenvalue $\omega$. We obtain the BLV state at reference point $x_0$ by integrating the dipole form $C(D_{x_1x_2})$ over its ends against the conjugate conformal eigenfunction,
\begin{equation}
\mathcal B_{n,\nu}(x_0)=\int\frac{d^2x_1\,d^2x_2}{|x_1-x_2|^4}\,E^{n,\nu}(x_1,x_2;x_0)^{*}\,C(D_{x_1x_2}).
\label{eq:blvform}
\end{equation}
The opposite labels $(n,\nu)$ and $(-n,-\nu)$ form one sector because they share the Casimir eigenvalues $h(1-h)$ and $\bar h(1-\bar h)$. By completeness of the conformal eigenfunctions, any superposition of dipoles at fixed labels is a superposition of these BLV states over $x_0$ (Appendix~\ref{sm:firstorder}).
A dipole with given ends contains a continuum of conformal labels. To use it in the eigenvalue equation one takes its component in this sector, retaining the integration over the reference point. Both terms with opposite labels have the same LO evolution.
The projectile component $\mathcal C_{n,\nu}$ is thus a superposition of BLV states over $x_0$. Its LO eigenvalue in the last line follows from Eq.~\eqref{eq:blveigen},
\begin{align}
\mathcal C_{n,\nu}(x,y)&=\frac{\nu^2+n^2/4}{\pi^4}\int d^2x_0\Big[E^{n,\nu}(x,y;x_0)\,\mathcal B_{n,\nu}(x_0)\nonumber\\
&\quad+E^{n,\nu}(x,y;x_0)^{*}\,\mathcal B_{-n,-\nu}(x_0)\Big],\nonumber\\
C(D_{xy})&=\sum_{n=1,3,\dots}\int d\nu\,\mathcal C_{n,\nu},\nonumber\\
H_0\,\mathcal C_{n,\nu}&=\bar\alpha_s\chi(n,\nu)\,\mathcal C_{n,\nu}.
\label{eq:Cright}
\end{align}
We perturb $H_0$ by a conformally invariant NLO term $V$ that respects permutation symmetry. The non-planar quark term has these properties. The terms proportional to the one-loop beta function coefficient $\beta_0=(11N_c-2n_f)/3$ break conformal invariance and thus fall outside this class of perturbations. We include the $\beta_0$ terms when we examine the color structures below. We denote the first-order corrections to the state and to the eigenvalue by $W^{(1)}_{n,\nu}$ and $\omega^{(1)}_V(n,\nu)$, respectively. Expanding the eigenvalue equation at the chosen labels yields
\begin{equation}
\big[H_0-\bar\alpha_s\chi(n,\nu)\big]\,W^{(1)}_{n,\nu}=\big[\omega^{(1)}_V(n,\nu)-V\big]\,\mathcal C_{n,\nu}.
\label{eq:firstordereq}
\end{equation}
We take $W^{(1)}_{n,\nu}$ and $V C(D_{xy})$ in the domain of Eq.~\eqref{eq:lefteigid}. Applying $\Pi_{n,\nu}$ then removes the unknown state correction because this overlap is a left eigenvector. The target integral commutes with the projectile Casimir operators, so it vanishes on a form at other labels. Conformal invariance makes $V$ diagonal in those labels. Thus the overlap with one insertion of $V$ selects the same projectile component as the overlap without an insertion, even if we use the full dipole (Appendix~\ref{sm:firstorder}, Eq.~\eqref{eq:labelsel}). The overlap with the unperturbed dipole is proportional to the normalization of cubic exchange and is nonzero for odd $n$ and real $\nu$ by Eqs.~\eqref{eq:projKq}, \eqref{eq:Z} and \eqref{eq:Zone}. Dividing by it gives
\begin{equation}
\omega^{(1)}_V(n,\nu)=\frac{\Pi_{n,\nu}\big[V\,C(D_{xy})\big]}{\Pi_{n,\nu}\big[C(D_{xy})\big]}.
\label{eq:firstorder}
\end{equation}
In the numerator of Eq.~\eqref{eq:firstorder} one insertion changes the state that scatters on the target. The denominator measures the overlap of the original projectile with that same target eigenfunction. Dividing removes the common normalization of the BLV state, the eigenfunction and the Born exchange. In particular, a term whose action has zero overlap at the branch point gives no first-order eigenvalue correction there.

Equation~\eqref{eq:firstorder} applies to an isolated BLV eigenvalue. At fixed conformal labels the LO Hamiltonian has discrete levels labeled by an integral of motion $q_3$~\cite{DerkachovKorchemskyKotanskiManashov:2002}. The BLV state at $(1,0)$ has $q_3=0$ and zero quasimomentum, the phase under cyclic permutation of the gluons. The non-planar quark term mixes the BLV state only with levels of zero quasimomentum (Appendix~\ref{sm:levels}). The BLV eigenvalue is isolated precisely when it differs from the eigenvalue of every other level of zero quasimomentum at the same conformal labels. We searched numerically for levels with $|q_3|<\nQthreeCutWide$ at $(1,0)$, and every level we found apart from the BLV state lay below the BLV eigenvalue. The nearest level of zero quasimomentum lay $\nGapBose\,\bar\alpha_s$ below it. The levels of zero quasimomentum we computed beyond the search lay further below the BLV eigenvalue (Appendix~\ref{sm:levels}).

The non-planar quark loop is the only candidate for a change of the overlap at the branch point in the NLO dipole equation~\cite{BalitskyChirilli:2008}. The color structures of Eq.~\eqref{eq:Bgluon} distinguish emission at one point $z$ from the two-gluon emission terms. For emission at one point the kernel $K_1$ of Eq.~\eqref{eq:K1} multiplies $S_1$. In the two-gluon emission terms the even kernel $K_2$ of Eq.~\eqref{eq:K2} multiplies $S_2$, and $K_3$ of Eq.~\eqref{eq:K3} multiplies $S_3$ (Appendix~\ref{sm:charge}). The quark kernel $K_q$ of Eq.~\eqref{eq:Kq} multiplies the color structure $S_q$ of Eq.~\eqref{eq:Sqquad}. In our notation the coefficient is $c_q=\alpha_s^2 n_f/(4\pi^4)$ and the dipole equation reads
\begin{align}
\frac{d}{dY}\,\mathrm{Tr}\{U_xU_y^\dagger\}={}&\frac{\alpha_s}{2\pi^2}\int d^2z\,K_1\,S_1\nonumber\\
&+\frac{\alpha_s^2}{16\pi^4}\int d^2z\,d^2z'\,\big[K_2\,S_2+K_3\,S_3\big]\nonumber\\
&+c_q\int d^2z\,d^2z'\,K_q\,S_q.
\label{eq:nloterms}
\end{align}
The overlap with the BLV state at the branch point is proportional to $L_g$ on superpositions of dipoles for the test function of Eq.~\eqref{eq:LgDipole}. A term that sends a dipole to a superposition of dipoles and conserves $L_g$ therefore leaves this overlap unchanged. The constants, the double logarithm and the $\beta_0$ terms all multiply the LO color structure $S_1$, so their cubic forms have both properties.
Each of these kernels weights a difference of dipole forms whose local coefficients already cancel at a given emission point. Changing the kernel multiplying this difference therefore leaves the cancellation intact, even for the terms with a scale in the kernel.
The two-gluon emission terms also reduce to dipoles. The functional $L_g$ vanishes pointwise for $S_2$ and after symmetric integration for $S_3$ (Appendix~\ref{sm:charge}).
In the latter case the local contributions at the parent ends remain before integration. They are the same for both orderings of the two emissions, whereas the kernel changes sign when the emission points are interchanged.
For the planar quark loop the cubic dipole contribution itself is odd under interchange of the emission points. Its integral against the even kernel $K_q$ converges absolutely and therefore vanishes in any integration order (Appendix~\ref{sm:charge}).

\begin{figure}[tp]
\centering
\includegraphics[width=\columnwidth]{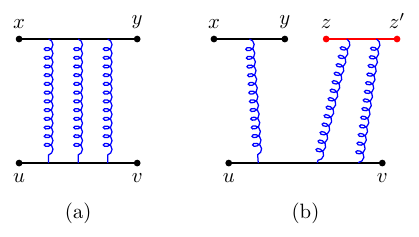}
\caption{Born exchange with the target dipole $(u,v)$. In (a) each of the three gluons couples to the projectile dipole $(x,y)$, giving $C(D_{xy})$ in Eq.~\eqref{eq:Cdipole}. Panel (b) shows the non-dipole state produced by the non-planar quark loop at distinct points. One gluon attaches to the parent and the other two to the dipole of the $q\bar q$ pair (red). These attachments give $T(D_{xy},D_{zz'},D_{zz'})$ in Eq.~\eqref{eq:quark}.}
\label{fig:states}
\end{figure}

For adjoint matter the color factor of the non-planar loop on a symmetric adjoint pair vanishes and the cubic forms are superpositions of dipoles. At an emission point a fundamental Wilson line and its conjugate representing an adjoint line have zero net sign. The virtual term completes the cancellation at the parent ends, so adjoint matter also conserves $L_g$. The non-planar loop of fundamental quarks is the remaining candidate for a first-order shift. In the state produced by this loop, one gluon attaches to the parent dipole and two attach to the dipole of the $q\bar q$ pair (Fig.~\ref{fig:states}).
Let $z$ and $z'$ be the quark and antiquark positions. In the bracket of the quark kernel, $\theta$ is the phase of the cross ratio of $x,y,z,z'$ and $\tau$ is the logarithm of its modulus. Substitution of the identity~\eqref{eq:Kqid} from Appendix~\ref{sm:color} yields
\begin{align}
K_q(x,y;z,z')&=\frac{4}{|z-z'|^4}\Big[1-\frac{\tau}{\sinh\tau}\,\cos\theta\Big],\nonumber\\
\tau+i\theta&=\ln\frac{(x-z)(y-z')}{(x-z')(y-z)}.
\label{eq:Kq}
\end{align}
Interchanging the emission points leaves the kernel unchanged, so only the even part of the cubic expansion of $S_q$ contributes. We normalize the fundamental generators by $\mathrm{tr}(t^at^b)=\delta^{ab}/2$ and subtract the second trace at coincident emission points. Two applications of the Fierz identity give
\begin{align}
S_q(z,z')={}&\mathrm{Tr}\{t^aU_xt^bU_y^\dagger\}\nonumber\\
&\times\big[\mathrm{Tr}\{t^aU_zt^bU_{z'}^\dagger\}-(z'\to z)\big]\nonumber\\
={}&\tfrac14\big[\mathrm{Tr}\{U_xU_{z'}^\dagger\}-\mathrm{Tr}\{U_xU_z^\dagger\}\big]\,\mathrm{Tr}\{U_zU_y^\dagger\}\nonumber\\
&-\frac{1}{4N_c}\Big[\mathrm{Tr}\{U_xU_y^\dagger U_zU_{z'}^\dagger\}+\mathrm{Tr}\{U_xU_{z'}^\dagger U_zU_y^\dagger\}\nonumber\\
&\qquad-\mathrm{Tr}\{U_xU_y^\dagger\}\Big]\nonumber\\
&+\frac{1}{4N_c^2}\,\mathrm{Tr}\{U_xU_y^\dagger\}\,\mathrm{Tr}\{U_zU_{z'}^\dagger\}.
\label{eq:Sqquad}
\end{align}
The two traces of four Wilson lines are quadrupoles. Each carries a factor of $1/N_c$, and the other terms are dipoles and products of dipoles. We denote the cubic imaginary part by an arrow, so the cubic $C$-odd part of a trace is $i$ times the expression on the right. Expanding the first quadrupole by Eq.~\eqref{eq:cubic} gives
\begin{align}
\mathrm{Tr}\{U_xU_y^\dagger U_zU_{z'}^\dagger\}&\to C(D_{xy}+D_{zz'})\nonumber\\
&=C(D_{xy})+3T(D_{xy},D_{xy},D_{zz'})\nonumber\\
&\quad+3T(D_{xy},D_{zz'},D_{zz'})+C(D_{zz'}).
\label{eq:quadcubic}
\end{align}
The second quadrupole has the same signed sum of fields $D_{xy}+D_{zz'}$ and hence the same cubic imaginary part.
The term with two gluons on the dipole of the $q\bar q$ pair has three contributions, one for each position of the parent argument. The symmetry of the trilinear form makes them equal, giving the coefficient of three in Eq.~\eqref{eq:quadcubic}.
The terms $C(D_{xy})$ and $3T(D_{xy},D_{zz'},D_{zz'})$ in each quadrupole are even under interchange of the emission points. The other traces in Eq.~\eqref{eq:Sqquad} cancel the contributions proportional to $C(D_{xy})$, and the planar dipole products together give only odd terms. Thus the entire even part of the cubic expansion of $S_q$ comes from the term with two gluons on the dipole of the $q\bar q$ pair in each quadrupole and is $-6\,T(D_{xy},D_{zz'},D_{zz'})/(4N_c)$.

The terms linear and cubic in $D_{zz'}$ in Eq.~\eqref{eq:quadcubic} are odd. The other traces cancel the term cubic in $D_{zz'}$ only in part, leaving $-C(D_{zz'})/(4N_c)$ in the cubic expansion of $S_q$ (Eq.~\eqref{eq:Bquark}). All odd terms of $S_q$ vanish upon integration against $K_q$. Integrating its even part gives the non-dipole state produced by the non-planar quark loop,
\begin{align}
\frac{d}{dY}\,C(D_{xy})\Big|_{q}=-\frac{6c_q}{4N_c}&\int d^2z\,d^2z'\,K_q(x,y;z,z')\nonumber\\
&\times T(D_{xy},D_{zz'},D_{zz'}),
\label{eq:quark}
\end{align}
where $D_{xy}$ describes the gluon on the parent and the two factors of $D_{zz'}$ the gluons on the $q\bar q$ dipole. At cubic order the quark part of the $C$-odd SU(3) equation of Ref.~\cite{BalitskyGrabovsky:2015} reduces to Eq.~\eqref{eq:quark} at $N_c=3$, including its coefficient (Appendix~\ref{sm:color}).

The linearized NLO dipole equation of Ref.~\cite{BalitskyChirilli:2008}, used in Ref.~\cite{Kovchegov:2013}, contains a two-gluon, $C$-even quark loop whose action on the dipole form $C(D_{xy})$ is odd under interchange of the $q\bar q$ positions. Integration against $K_q$ thus gives
\begin{align}
\frac{d}{dY}\,C(D_{xy})\Big|_{q,\,{\rm lin}}&=-\frac{c_q}{4N_c}\int d^2z\,d^2z'\,K_q(x,y;z,z')\nonumber\\
&\quad\times C(D_{zz'})=0.
\label{eq:qlin}
\end{align}
The gluon terms and the planar quark loop agree in the two calculations. The full structure $S_q$ also produces the non-dipole state of Eq.~\eqref{eq:quark}. At distinct ends its cubic form contains products of fields at three different points, whereas a dipole form involves only its two ends.
A gluon on the parent can coincide with neither end of the dipole of the $q\bar q$ pair in this configuration. The two gluons on that dipole can be at the same end or at opposite ends, but in both cases there is no local term with all three gluons at one point.
At each point at least one of the three arguments therefore has zero net sign, so the functional $L_g$ vanishes. The proportionality in Eq.~\eqref{eq:Pi0dipole} holds on superpositions of dipoles and does not apply to this non-dipole state.

We evaluate its overlap with the BLV state to determine the sign and size of its contribution. We denote the two-gluon exchange of the $q\bar q$ dipole integrated against $K_q$ by $R_q$. The parent contributes one factor of $l$. The integration measure cancels the conformal factors of $K_q$, so the non-planar quark term is conformally invariant. Equation~\eqref{eq:firstorder} thus determines the non-planar quark contribution $\delta\omega(\nu)$ to the $n=1$ eigenvalue,
\begin{align}
\delta\omega(\nu)&=-\frac{6c_q}{4N_c}\,
\frac{\Pi_{1,\nu}\big[l(x,y;u,v)\,R_q(x,y;u,v)\big]}{\Pi_{1,\nu}\big[l(x,y;u,v)^3\big]},\nonumber\\
R_q(x,y;u,v)&=\int d^2z\,d^2z'\,K_q(x,y;z,z')\,l(z,z';u,v)^2.
\label{eq:ratio}
\end{align}
Both $lR_q$ and $l^3$ depend on the four dipole ends through their cross ratio $\rho$. For these two amplitudes $F$, the target integration at $(1,0)$ reduces to an inner product on the cross-ratio plane. We derive the reduction in Appendix~\ref{sm:pairs},
\begin{align}
\Pi_{1,0}[F]&=2\pi\,\langle l|F\rangle\,E^{1,0}(x,y;x_0),\qquad
\langle f|g\rangle=\int\frac{d^2\rho}{|\rho|^4}\,f\,g,\nonumber\\
\rho&=\frac{(x-y)(u-v)}{(x-u)(y-v)}.
\label{eq:Pi0rho}
\end{align}
The factor $l$ from the target integration combines with the one from the parent dipole. The common dependence on $x,y$ cancels in the ratio.
For Eq.~\eqref{eq:main} we need the convergent overlaps at $(1,0)$,
\begin{equation}
\delta\omega(0)=-\frac{6c_q}{4N_c}\,\frac{\langle l^2|R_q\rangle}{\langle l^2|l^2\rangle},\qquad
\langle l^2|l^2\rangle=3\pi\zeta(3).
\label{eq:expect}
\end{equation}
The numerator now compares the two-gluon exchange integrated over the $q\bar q$ positions with the Born exchange of two gluons. The denominator is the norm of that same Born exchange. The measure on the cross-ratio plane in Eq.~\eqref{eq:Pi0rho} is positive. Since the Born exchange $l^2$ is nonnegative, a positive $R_q$ for separated dipoles gives a positive numerator without evaluating its integral.
The kernel fixes the sign of the overlap. For real $\tau$ the ratio $\tau/\sinh\tau$ is positive and at most one. The bracket in Eq.~\eqref{eq:Kq} is consequently positive except on a set of measure zero. Thus $R_q$ for separated dipoles and its overlap with $l^2$ are positive. The coefficient $-6c_q/(4N_c)$ in Eq.~\eqref{eq:quark} is negative because $c_q$ is positive for $n_f>0$. This negative coefficient and the positive overlap imply $\delta\omega(0)<0$.

We have obtained Eq.~\eqref{eq:firstorder} by projecting the eigenvalue equation with a left eigenvector that also acts on non-dipole forms. Only the non-planar quark loop in Eq.~\eqref{eq:nloterms} changes the overlap at the branch point. For the state of Eq.~\eqref{eq:quark}, the ratio of overlaps in Eq.~\eqref{eq:expect} is positive. Its color coefficient therefore gives a negative first-order correction at the branch point if the BLV eigenvalue there is isolated. In the next section we determine the size of the overlap by resolving one gluon pair into conformal partial waves.

\section{Conformal partial waves of a gluon pair in the BLV state}
\label{sec:result}
We compute the size of $\delta\omega(0)$ at the branch point $(1,0)$ from the two-gluon exchange in Eq.~\eqref{eq:expect}. We need its distribution over conformal partial waves because the quark kernel acts with a different eigenvalue on each pair of labels. From here on $(n',\nu')$ labels a gluon pair and $(n,\nu)$ labels the three-gluon state. Interchanging the dipole ends reverses $l$ but preserves $l^2$. The pair therefore has even conformal spin $n'$, although the whole BLV state at the branch point has $n=1$.

\begin{figure}[tp]
\centering
\includegraphics[width=\columnwidth]{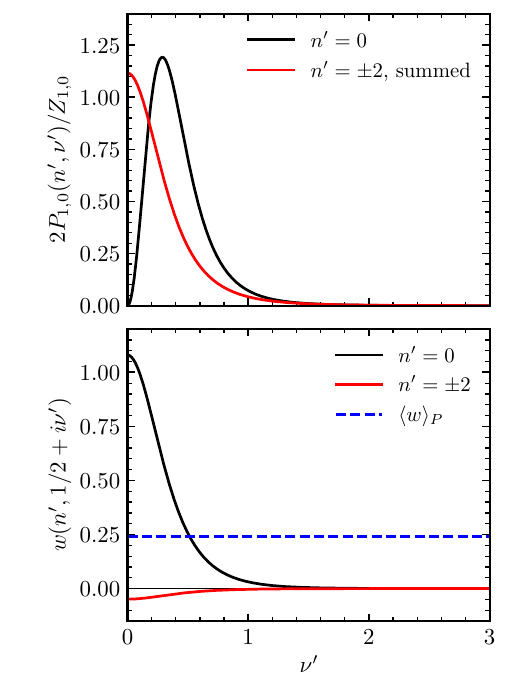}
\caption{Distribution of the norm of two-gluon exchange at $(n,\nu)=(1,0)$. The upper curves show twice $P_{1,0}$ divided by the norm $Z_{1,0}=\langle l^2|l^2\rangle=3\pi\zeta(3)$. Their areas over $\nu'\geq0$ thus give the fractions of the norm carried by $n'=0$ (black) and $n'=\pm2$ together (red), including both signs of $\nu'$. The conformal partial waves with even $|n'|\geq4$ carry less than \nHighSpinPercentWord{} percent of the norm. The lower panel shows $w(n',1/2+i\nu')$ for the same conformal spins. The dashed blue line marks the average over the distribution, Eq.~\eqref{eq:wP}.}
\label{fig:pairs}
\end{figure}

We write the non-planar quark term of the NLO BFKL eigenvalue of Kotikov and Lipatov~\cite{KotikovLipatov:2000} for a color-singlet pair through the function $w$. Its arguments are the conformal spin and the Mellin label $\gamma$ of the pair. In terms of general arguments the function reads
\begin{align}
w(n,\gamma)&=\frac{\cos\pi\gamma}{(1-2\gamma)\sin^2\pi\gamma}\nonumber\\
&\quad\times\frac{\big[2+3\gamma(1-\gamma)\big]\delta_{n,0}-\tfrac12\gamma(1-\gamma)\,\delta_{|n|,2}}{(3-2\gamma)(1+2\gamma)}.
\label{eq:w}
\end{align}
To use this eigenvalue in dipole scattering we combine the conformal eigenfunction on the exchanged gluon pair with its conjugate on the target.
The quark kernel acts diagonally on the conformal partial waves. The eigenfunction and its conjugate have the same spin and real label in the completeness relation~\eqref{eq:completeness} of Appendix~\ref{sm:firstorder}, which we use to expand the delta functions for Eq.~\eqref{eq:l2expand} in Appendix~\ref{sm:pairs}. Integration over the common reference point leaves a function of the two gluon positions and the two target ends,
the conformal partial wave
\begin{align}
\Psi_{n',\nu'}(z,z';u,v)&=\int d^2x_0\,E^{n',\nu'}(z,z';x_0)\nonumber\\
&\quad\times E^{n',\nu'}(u,v;x_0)^{*}.
\label{eq:Phidef}
\end{align}
The integral of $K_q$ over the $q\bar q$ positions multiplies each partial wave by its pair eigenvalue. The quadratic dipole term~\eqref{eq:Bsinglet} describes the same two-gluon exchange in a color singlet. Matching its coefficient to the non-planar quark term of the color-singlet NLO BFKL eigenvalue in Appendix~\ref{sm:color} fixes the normalization,
\begin{widetext}
\begin{equation}
\int d^2z\,d^2z'\,K_q(x,y;z,z')\,\Psi_{n',\nu'}(z,z';u,v)=4\pi^4\,w\big(n',\tfrac12+i\nu'\big)\,\Psi_{n',\nu'}(x,y;u,v)\qquad(n'\ \text{even}).
\label{eq:Kqw}
\end{equation}
\end{widetext}
The $q\bar q$ integration can now be done independently for each pair of labels. To combine the results for a physical dipole we need the coefficient of each partial wave in its exchange $l^2$. We recover the Born coefficient by dividing by the eigenvalue of the conformal Laplacian on that wave. The denominator $\mathcal D_1$ depends on the conformal labels $(n',\nu')$ of the conformal partial wave,
\begin{equation}
\mathcal{D}_1(n',\nu')=\big[\nu'^2+\tfrac{(n'-1)^2}{4}\big]\big[\nu'^2+\tfrac{(n'+1)^2}{4}\big].
\label{eq:Ddef}
\end{equation}
Acting on the Born exchange with a Laplacian at each gluon position gives delta functions that place the gluons at the two target ends.
The two assignments to the target ends enter with the same sign at distinct gluon positions, so their sum contains only conformal partial waves of even spin.
The expansion of these delta functions by completeness gives the Born exchange in conformal partial waves, with the denominator of Eq.~\eqref{eq:Ddef} in each coefficient (Appendix~\ref{sm:pairs}),
\begin{align}
l(z,z';u,v)^2&=\sum_{n'\,\mathrm{even}}\int d\nu'\,\frac{\nu'^2+n'^2/4}{\pi^4}\nonumber\\
&\quad\times\frac{\pi^2}{\mathcal{D}_1(n',\nu')}\,\Psi_{n',\nu'}(z,z';u,v).
\label{eq:l2expand}
\end{align}
Each term of the Born expansion enters Eq.~\eqref{eq:expect} through the overlap of its partial wave with $l^2$. The numerator contains one extra factor, the eigenvalue in Eq.~\eqref{eq:Kqw}. To determine how much each wave contributes to the average we evaluate the overlap $\Pi_{1,0}$ of $l$ times one conformal partial wave with the BLV state. After removal of the common factor $2\pi E^{1,0}$ this equals the inner product of the partial wave with $l^2$ (Appendix~\ref{sm:pairs}). Multiplying this inner product by the coefficient of the same wave in the Born expansion gives the nonnegative distribution $P_{1,0}(n',\nu')$.
The integral over an interval of real $\nu'$ at fixed even spin gives the share of the norm of the Born exchange carried by the conformal partial waves of one gluon pair in that interval.
The contributions of the waves $\Psi_{n',\nu'}$ integrate and sum to the norm $\langle l^2|l^2\rangle$ of two-gluon exchange. Dividing by the integral of $P_{1,0}$ normalizes the average $\langle f\rangle_P$ of a function of the pair labels,
\begin{align}
\langle l^2|\Psi_{n',\nu'}\rangle&=\frac{\pi^2}{\mathcal{D}_1(n',\nu')},\nonumber\\
P_{1,0}(n',\nu')&=\frac{\nu'^2+n'^2/4}{\mathcal{D}_1(n',\nu')^2},\nonumber\\
\langle f\rangle_P&=\frac{\sum_{n'\,\mathrm{even}}\int d\nu'\,P_{1,0}(n',\nu')\,f(n',\nu')}{\sum_{n'\,\mathrm{even}}\int d\nu'\,P_{1,0}(n',\nu')}.
\label{eq:P10}
\end{align}
Each of the three gluon pairs is a symmetric adjoint pair for every $N_c\geq3$. The non-planar loop has color factor $-1/(2N_c)$ on such a pair and $-1/(4N_c)$ on a color-singlet pair (Eq.~\eqref{eq:Bcross}). The symmetric and antisymmetric adjoint pair kernels of Ref.~\cite{BartelsFadinLipatovVacca:2013} differ by the Abelian part of the symmetric kernel. The loop in Eq.~\eqref{eq:quark} gives this part and hence twice the singlet eigenvalue on $\Psi_{n',\nu'}$.
Permutation symmetry gives the same color projection and the same distribution of the two-gluon exchange for each choice of spectator. The factor of three in Eq.~\eqref{eq:avg10} counts these choices. The factor of two multiplies the non-planar quark term of Kotikov and Lipatov on a color-singlet pair, with the average $\langle w\rangle_P$ over the conformal labels of one gluon pair.
Inserting the eigenvalue of Eq.~\eqref{eq:Kqw}, written through $w$ in Eq.~\eqref{eq:w}, in the numerator gives
\begin{align}
\langle l^2|R_q\rangle&=4\pi^4\sum_{n'\,\mathrm{even}}\int d\nu'\,P_{1,0}(n',\nu')\,w\big(n',\tfrac12+i\nu'\big),\nonumber\\
\langle l^2|l^2\rangle&=\sum_{n'\,\mathrm{even}}\int d\nu'\,P_{1,0}(n',\nu')=3\pi\zeta(3),\nonumber\\
\delta\omega(0)&=3\times2\times\Big(-\frac{n_f\alpha_s^2}{4N_c}\Big)\langle w\rangle_P\nonumber\\
&=-\frac{3n_f\alpha_s^2}{2N_c}\,\langle w\rangle_P.
\label{eq:avg10}
\end{align}
The average lies below the largest value of $w$ (Fig.~\ref{fig:pairs}). We find normalized areas $\nAreaZero$ for $n'=0$ and $\nAreaTwo$ for $n'=\pm2$ together. The positive maximum $w(0,1/2)=\nWPomeron$ determines the non-planar quark term of the NLO pomeron intercept. At this point the $n'=0$ distribution vanishes quadratically in $\nu'$.
The $n'=0$ contribution therefore samples the function on either side of its maximum. Even spins above two in absolute value enter the norm in the denominator, although the quark eigenvalue vanishes there.
Including the small negative contribution from $n'=\pm2$ brings the average to about $\nWAvgShort$. The residue integrals in Appendix~\ref{sm:closed} give its analytic value,
\begin{align}
\langle w\rangle_P&=\frac{91\pi^4+90\pi^2-1900+4260\,\zeta(3)+5400\,\zeta(5)}{6480\,\pi^2\,\zeta(3)}\nonumber\\
&=\nWAvg.
\label{eq:wP}
\end{align}
Inserted in Eq.~\eqref{eq:avg10}, this average gives Eq.~\eqref{eq:main} when the maximum lies at $\nu=0$ (Sec.~\ref{sec:location}). The form of Eq.~\eqref{eq:quark} is non-dipole even after integration over the $q\bar q$ positions because $L_g$ still vanishes while the overlap with the BLV state is proportional to the nonzero average in Eq.~\eqref{eq:avg10}.
To locate the maximum in Sec.~\ref{sec:location} we also need the non-planar quark contribution away from the branch point along $n=1$. We compute the overlap for general $(n,\nu)$ using the difference between the Casimir eigenvalues of the full state and of one pair. The squared modulus of that difference factors into $\mathcal D_-\mathcal D_+$, with
\begin{equation}
\mathcal{D}_\mp=(\nu\mp\nu')^2+\tfrac{(n\mp n')^2}{4}.
\label{eq:Dmp}
\end{equation}
The spectator contributes the factor $l$, which changes sign under interchange of the target ends. The overlap of $l$ times one conformal partial wave with the BLV state therefore vanishes for even $n+n'$, when the integrand is odd under this interchange.
For the other parity the Casimir eigenvalue difference determines the denominator. Both factors are positive for real labels when the two spins have opposite parity, including at the branch point. This gives the finite overlap needed for each even-spin pair in the odd-spin state.
We obtain the overlap for all real labels by the calculation in Appendix~\ref{sm:pairs},
\begin{align}
\Pi_{n,\nu}\big[l\,\Psi_{n',\nu'}\big]&=\frac{2\pi^3}{\mathcal{D}_-\mathcal{D}_+}\,E^{n,\nu}(x,y;x_0)\qquad(n+n'\ \text{odd}),\nonumber\\
\Pi_{n,\nu}\big[l\,\Psi_{n',\nu'}\big]&=0\qquad(n+n'\ \text{even}).
\label{eq:single}
\end{align}
We multiply the overlap $\Pi_{n,\nu}$ in Eq.~\eqref{eq:single} by the Born coefficient in Eq.~\eqref{eq:l2expand} and remove the common factor $2\pi E^{n,\nu}$, just as at the branch point. The resulting distribution $P_{n,\nu}$ has integral $Z_{n,\nu}$, including the sum over even pair spins in Eq.~\eqref{eq:Z}. Its dependence on the conformal labels is
\begin{align}
P_{n,\nu}(n',\nu')&=\frac{\nu'^2+n'^2/4}{\big[\nu'^2+\frac{(n'-1)^2}{4}\big]\big[\nu'^2+\frac{(n'+1)^2}{4}\big]}\nonumber\\
&\times\frac{1}{\big[(\nu-\nu')^2+\frac{(n-n')^2}{4}\big]\big[(\nu+\nu')^2+\frac{(n+n')^2}{4}\big]}.
\label{eq:P}
\end{align}
We use $P_{n,\nu}$ in place of $P_{1,0}$ in the normalized average of Eq.~\eqref{eq:P10}. The projectile eigenfunction cancels between the numerator and denominator (Appendix~\ref{sm:pairs}), so for an isolated state the first-order non-planar quark contribution is
\begin{equation}
\delta\omega(n,\nu)=-\frac{3n_f\alpha_s^2}{2N_c}\,\langle w\rangle_P.
\label{eq:avg}
\end{equation}
At LO one symmetric adjoint pair has half the color-singlet eigenvalue. Three times the average of this symmetric adjoint eigenvalue over $P_{n,\nu}$ gives the LO eigenvalue of the BLV state.
These two relations give the LO identity for the distribution $P_{n,\nu}$ derived in Appendix~\ref{sm:pairs},
\begin{equation}
\big\langle\chi(n',\nu')\big\rangle_P=\tfrac23\,\chi(n,\nu).
\label{eq:LOid}
\end{equation}
The distribution $P_{n,\nu}$ is a property of the BLV state and its Born exchange. Since the action of the LO kernel on the full state is known, Eq.~\eqref{eq:LOid} tests how that exchange is distributed over the conformal partial waves of one gluon pair.
The argument of Sta\'sto~\cite{Stasto:2009} uses the color-singlet BFKL eigenvalue at $n=1$, for which the non-planar quark term vanishes by Eq.~\eqref{eq:w}. The BLV state also has total spin $n=1$, but the loop acts on its even-spin gluon pairs and the same term is nonzero for $n'=0,\pm2$.

We have evaluated the average $\langle w\rangle_P$ at the branch point in Eq.~\eqref{eq:wP}. Each of the three gluon pairs contributes twice the non-planar quark term on a color-singlet pair with this average. Their sum gives the coefficient in Eq.~\eqref{eq:main} once the maximum of the real part of the eigenvalue lies at $\nu=0$. Equation~\eqref{eq:avg} gives the first-order non-planar quark contribution for an isolated state at general labels. In the next section we use that dependence on the labels to locate the cut.

\section{The location of the branch point}
\label{sec:location}
We use the non-planar quark contribution at general labels to locate the cut. We must now combine that contribution with the other terms in the eigenvalue before taking its maximum. The dipole couples to a superposition over $\nu$ of BLV states with the LO partial-wave coefficients of Kovchegov, Szymanowski and Wallon~\cite{KovchegovSzymanowskiWallon:2004}. Near $\nu=0$ these coefficients have a regular expansion with a nonzero constant term (Appendix~\ref{sm:setting}, Eq.~\eqref{eq:Zone}). Each state evolves with $\exp[Y\omega(\nu)]$.
As rapidity increases, states with a smaller real part of the eigenvalue fall off relative to those near its maximum. The nonzero LO partial-wave coefficients near $\nu=0$ allow states in this region to contribute to the dipole amplitude. The integral over real labels gives a cut whose rightmost point is set by the largest real part of the eigenvalue in the superposition.
We therefore maximize the real part of the $n=1$ eigenvalue over real $\nu$. Its non-planar quark contribution follows from Eq.~\eqref{eq:avg} at this spin, with the average given in closed form by Eq.~\eqref{eq:Fnu}.

The composite dipole operator of Ref.~\cite{BalitskyChirilli:2009} differs from the Wilson-line dipole by a gluon term of order $\alpha_s$ with the LO color structure $S_1$. We use the composite operator for the terms of Sec.~\ref{sec:shift} that send a dipole to a superposition of dipoles because its linearized equation is the NLO BFKL equation at odd conformal spin~\cite{Kovchegov:2013}.
The planar NLO term of its $n=1$ eigenvalue is therefore the color-singlet NLO eigenvalue of Kotikov and Lipatov~\cite{KotikovLipatov:2000} at this spin with the $\beta_0$ terms and the non-planar quark term removed (Appendix~\ref{sm:branch}). The change of operator leaves the quark term unchanged through NLO. For the composite dipole we can therefore use the non-planar quark contribution of Eq.~\eqref{eq:avg} with the same average over the conformal labels of one gluon pair.
We take the $\beta_0$ terms from the color-singlet NLO eigenvalue of Kotikov and Lipatov (Eqs.~\eqref{eq:omegab} and~\eqref{eq:omegamu}). The renormalization scale is $\mu$, and $q,q'$ are the transverse gluon momenta at the projectile and target ends. The planar NLO term involves the function $\Phi$ of Ref.~\cite{KotikovLipatov:2000}, given at $n=1$ in Eq.~\eqref{eq:PhiKL}, and equals
\begin{align}
&\frac{\bar\alpha_s^2}{4}\Big[\Big(\frac{67}{9}-2\zeta(2)-\frac{10n_f}{9N_c}\Big)\chi(1,\nu)+6\zeta(3)\nonumber\\
&\quad+2\,\mathrm{Re}\,\psi''(1+i\nu)-4\,\mathrm{Re}\,\Phi\big(1,\tfrac12+i\nu\big)\Big].
\label{eq:planar}
\end{align}
We add the planar term, the $\beta_0$ terms and the non-planar quark contribution to the LO eigenvalue. The real part is even on the line $\gamma=1/2+i\nu$.
The change of dipole operator contributes an imaginary, odd term on this line at zero momentum transfer. Thus the real part used to locate the cut is the same for the composite and Wilson-line dipoles through NLO (Appendix~\ref{sm:branch}).

The $\beta_0$ terms are evaluated at fixed renormalization scale $\mu^2$ and fixed external scales $q^2,q'^2$. We use the $\overline{\rm MS}$ coupling and choose the symmetric energy scale $s_0=|q||q'|$~\cite{KotikovLipatov:2000}. The reference scale is $\mu^2=q^2$. For the numerical example we take $\alpha_s(q^2)=\nAlphaRef$ and $N_c=n_f=3$.

Along $n=1$ the average in Eq.~\eqref{eq:avg} has the closed form of Eq.~\eqref{eq:Fnu}. In units of $n_f\alpha_s^2/N_c$ the non-planar quark contribution on this line is $-(3/2)\langle w\rangle_P(\nu)$. We denote the dimensionless coefficient of $\nu^2$ in this function by $a_q$. Expansion of the closed form about $\nu=0$, in the notation of Eq.~\eqref{eq:Fexp}, gives $a_q\approx\nCurvQuark$. This is the non-planar quark contribution to the quadratic coefficient of the real part of the eigenvalue in units of $n_f\alpha_s^2/N_c$ (Eq.~\eqref{eq:omegaexp}). Its positive sign reduces the magnitude of the negative LO quadratic coefficient at weak coupling.
At the reference scale the expansion about $\nu=0$ gives
\begin{widetext}
\begin{align}
\mathrm{Re}\,\omega(\nu)={}&\delta\omega(0)-\bigg\{2\zeta(3)\,\bar\alpha_s-\frac{\bar\alpha_s^2}{4}\Big[35\zeta(5)-2\zeta(3)\Big(\frac{67}{9}-2\zeta(2)-\frac{10n_f}{9N_c}\Big)\Big]\nonumber\\
&-a_q\,\frac{n_f\alpha_s^2}{N_c}\bigg\}\,\nu^2+O(\nu^4).
\label{eq:omegaexp}
\end{align}
\end{widetext}
The negative LO curvature dominates at weak coupling. At first order a displacement of the maximum is proportional to the slope of the NLO correction at $\nu=0$. Evenness makes the slope vanish, so the maximum stays at $\nu=0$.
\begin{figure}[tp]
\centering
\includegraphics[width=\columnwidth]{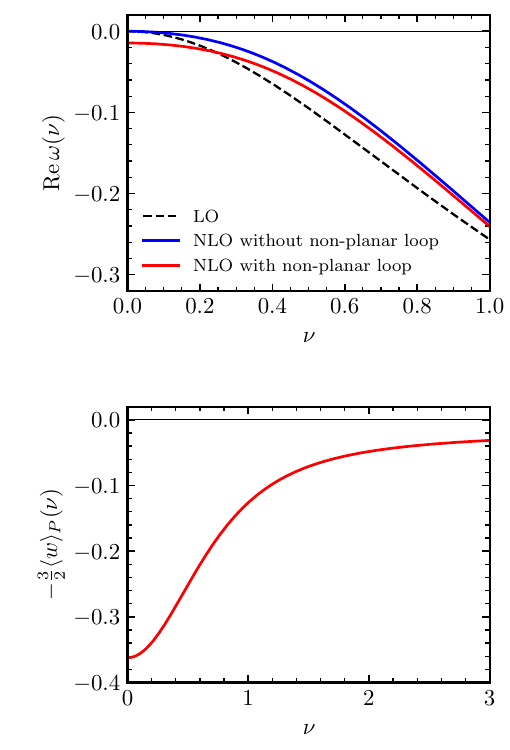}
\caption{Real part of the $n=1$ BLV eigenvalue (upper panel, $0\leq\nu\leq1$). The $\overline{\rm MS}$ coupling is $\alpha_s(q^2)=\nAlphaRef$ at $\mu^2=q^2$, with $N_c=n_f=3$ and $s_0=|q||q'|$ for projectile and target gluon momenta $q,q'$. LO (dashed) denotes $\bar\alpha_s\chi(1,\nu)$. NLO without the non-planar loop (blue) adds the planar NLO term and the real $\beta_0$ terms. NLO with the non-planar loop (red) adds the first-order non-planar quark contribution $\delta\omega(\nu)$. The lower panel extends the difference of the upper curves (red minus blue), divided by $n_f\alpha_s^2/N_c$, to $\nu=\nNuEdge$. It equals $-(3/2)\langle w\rangle_P(\nu)$ from Eq.~\eqref{eq:Fnu}. The first-order non-planar quark contribution assumes that the BLV state stays isolated along the curve (Sec.~\ref{sec:location}). We computed the curves in both panels at $\nCurveDigits$ digits, using the integral for $\Phi$ in Eq.~\eqref{eq:PhiKL} for the planar NLO term in the upper panel (Appendix~\ref{sm:methods}).}
\label{fig:shift}
\end{figure}
Its value is $\delta\omega(0)$ because the LO term, the planar NLO term and the $\beta_0$ terms vanish there, giving the branch point~\eqref{eq:main}.
The sign of the net quadratic coefficient determines whether $\nu=0$ is a local maximum of the real part. At the reference scale both the planar NLO term and the non-planar quark term contribute to this coefficient in Eq.~\eqref{eq:omegaexp}.

For the reference parameters we find the maximum numerically at $\nu=0$ (Fig.~\ref{fig:shift}). The NLO curves at nonzero $\nu$ use the $\beta_0$ terms at the specified fixed scales.
The full NLO curve (red) includes the first-order non-planar quark contribution of Eq.~\eqref{eq:avg}. At $\nu=0$ its difference from the curve without this term is the shift of the branch point in Eq.~\eqref{eq:main}. The lower panel gives the size of the non-planar quark contribution along the curve in units of $n_f\alpha_s^2/N_c$.
We use the first-order non-planar quark contribution for $0\leq\nu\leq\nNuEdge$ because no other level of zero quasimomentum reached the BLV eigenvalue among the levels found and followed numerically (Appendix~\ref{sm:levels}). The gap to the nearest level of zero quasimomentum decreased from $\nGapBose\,\bar\alpha_s$ at $\nu=0$ to $\nGapNuOne\,\bar\alpha_s$ at $\nu=1$ and $\nGapNuThree\,\bar\alpha_s$ at $\nu=\nNuEdge$.

The reference value of $\alpha_O-1$ is $-\nShiftCentral$. To vary the scale we hold $\alpha_s(q^2)=\nAlphaRef$ fixed at three colors and three flavors and convert to $\alpha_s(\mu^2)$ at one loop. The value at $\nu=0$ then changes by terms of order $\alpha_s^3$.
Away from $\nu=0$ the explicit scale logarithm multiplies the LO characteristic function. It contributes to the quadratic coefficient and cancels the change of the LO eigenvalue through NLO. The remaining dependence of the curvature on the renormalization scale can change the position of the maximum in the fixed-scale evaluation, since the NLO terms already reduce the magnitude of the LO curvature at the reference scale.
The full quadratic coefficient evaluated with the fixed-scale $\beta_0$ terms changes sign at about $\mu^2=\nMuThreshold\,q^2$, and the maximum lies at $\nu=0$ from that scale up to $4q^2$. At $q^2/4$ the curvature at $\nu=0$ is positive. The full eigenvalue then has a displaced maximum near $\nu=\nNuMaxLow$ (Appendix~\ref{sm:branch}). The shift ranges from $-\nShiftMuLowMax$ at the displaced maximum for $q^2/4$ to $-\nShiftMuHigh$ at $\nu=0$ for $4q^2$.
Each end of this scale band uses the maximum of the curve at its own renormalization scale.

The maximum determines the leading large-rapidity behavior of the $n=1$ amplitude at zero momentum transfer for oriented dipoles.
The overlap of cubic exchange at an infinite reference point and its normalization determine the $|n|=1$ part of the LO amplitude integrated over the relative impact parameter (Appendix~\ref{sm:pairs}).
In the integral over $\nu$ the coefficient $2\pi Z_{\pm1,\nu}$ multiplies the LO evolution factor and a power of the ratio of the dipole sizes.
Adding the two spins gives the cosine of the angle between the dipole axes, which factors out of the integral together with the product of the dipole sizes and a constant.
The exponent depends on the integration variable, and the power of the size ratio equals one at $\nu=0$.
There $2\pi Z_{\pm1,\nu}$ is regular and nonzero, and its value times the factors independent of $\nu$ gives the coefficient of the large-rapidity fall.
Since the maximum is quadratic, the LO amplitude falls like $(\bar\alpha_s Y)^{-1/2}$ for dipoles of fixed sizes and orientations whose axes are not perpendicular.
At LO the evolution factor is unity at the maximum itself. The fall comes from integration over a narrowing interval of labels on either side. After the intercept is lowered, the evolution factor at the maximum also decreases with rapidity.
At weak fixed coupling the shift multiplies the LO behavior by $\exp[-\nShiftCoeffShort\,n_f\alpha_s^2Y/N_c]$, so the BLV contribution falls faster with increasing energy.

We have located the weak-coupling maximum at $\nu=0$ using the evenness of the real part and its curvature in Eq.~\eqref{eq:omegaexp}. The scale band compares the positions of the branch point at fixed reference coupling, including the displaced maximum at the lower scale. The intercept fixes the exponential factor in the rapidity dependence of the cut's contribution. In the Conclusions we discuss what this fall implies for the spin-dependent amplitude and how to test the effect of the quark loop with running coupling.

\section{Conclusions}
\label{sec:concl}
We calculated the shift of the BLV branch point at zero momentum transfer to first order in the NLO kernel. We kept dilute three-gluon exchange at cubic order in the target field without saturation.
The calculation uses QCD with $N_c\geq3$ colors and $n_f$ massless fundamental flavors, where $n_f<11N_c/2$.
At weak fixed coupling the real part of the $n=1$ eigenvalue reaches its maximum at $\nu=0$. We have shown that non-planar quark loops reduce the BLV intercept by $\nShiftCoeff\,n_f\alpha_s^2/N_c$ if the BLV state at $(1,0)$ is isolated (Eq.~\eqref{eq:main}). The entire reduction comes from the non-dipole state, in which two exchanged gluons attach to the $q\bar q$ dipole and one to the parent.
The BLV intercept for gluons stays at one even with adjoint matter that keeps the theory asymptotically free.

At fixed $\bar\alpha_s$ the shift vanishes in the planar limit, in agreement with the planar results of Caron-Huot~\cite{CaronHuot:2015} and of Bartels and Vacca~\cite{BartelsVacca:2013} under the generalized bootstrap assumption. At fixed $\bar\alpha_s$ and fixed $n_f/N_c$, the shift falls as $1/N_c^2$.

The scale variation at fixed reference coupling is an example of the order-$\alpha_s^3$ corrections beyond the calculation. The NLO curves at $\nu\ne0$, the scale variation of the $\nu^2$ coefficient with its sign change near $\nMuThreshold\,q^2$, and the displaced maximum all use the fixed-scale $\beta_0$ terms of Kotikov and Lipatov. The action of the $\beta_0$ terms on the conformal partial waves away from $\nu=0$ remains to be derived. Running of the coupling during evolution and states of five or more reggeized gluons also lie outside our calculation. Such states enter the full QCD odderon intercept and can have intercept one at LO in the planar limit~\cite{DerkachovKorchemskyKotanskiManashov:2002}.

The $C$-odd amplitude on a transversely polarized proton enters the leading eikonal contribution to the quark Sivers function at small Bjorken $x_B$~\cite{KovchegovSantiago:2022}. The factor $x_B^{1-\alpha_O}$ would suppress the leading term relative to an intercept of one wherever the BLV cut governs that term at fixed coupling, with stronger suppression as $x_B$ decreases.

We expect the non-planar loop to suppress the amplitude by a scale-dependent factor even if the coupling runs during evolution. At large transverse gluon momentum $k_\perp$ we expect the LO eigenvalue to become independent of $\nu$ and the non-planar quark contribution to tend to zero, so the branch point could return to one. A numerical LO calculation at zero momentum transfer found a cut starting at one with running coupling and an infrared cutoff~\cite{BartelsContrerasVacca:2020}. We propose testing this expectation by adding the non-planar quark term proportional to $\alpha_s(k_\perp^2)^2$ to the $n=1$ eigenvalue equation of Ref.~\cite{BartelsContrerasVacca:2020} with the same running coupling and infrared cutoff.

The distribution $P_{n,\nu}$ also permits a first-order calculation for an isolated BLV state whenever the same conformally invariant operator acts on every pair. The eigenvalues of that operator on one gluon pair suffice.
We trace the shift of the branch point entirely to the non-dipole state of the non-planar quark loop.
At fixed coupling the lowering by non-planar quark loops suppresses the contribution of the BLV cut to the dipole amplitude by a factor that decreases exponentially with rapidity.

\section*{Data availability}
The programs and numerical data that support the findings of this article are openly available in the ancillary files of its arXiv version.

\begin{acknowledgments}
We thank Sergey Bondarenko for fruitful discussions. This research is supported in part by the Council of Higher Education of Israel through the grant ``Support Program of the High Energy Physics''.

Statement on the use of AI tools. During the preparation of this work, the authors used Claude Fable 5.1, Claude Fable 5.2, Claude Opus 5.5, GPT-5.6 Sol, GPT-6.1 Sol, GPT-6.2 Sol, GPT-6 Astra, GPT-6.2 Astra, and Gemini 4 Carbon to write symbolic, numerical, and plotting code, to carry out and check parts of the derivations, and to draft and edit the text. After using these tools, the authors reviewed and verified the code, the derivations, the outputs and the text, and take full responsibility for the content of the publication.
\end{acknowledgments}

\appendix

\section{The leading-order spectrum}
\label{sm:setting}
We determine how the dipole couples to the LO spectrum near the maximum used to locate the branch point in Sec.~\ref{sec:location}. The maximum of the LO characteristic function~\eqref{eq:chi} at odd conformal spin is zero. It occurs at $|n|=1$ and $\nu=0$, with $\chi(1,\nu)$ starting at $-2\zeta(3)\nu^2$. Our $\chi$ is twice the characteristic function used by Kovchegov, Szymanowski and Wallon~\cite{KovchegovSzymanowskiWallon:2004}. The eigenvalue of $K_0$ is then $2\pi\chi$, corresponding to $\bar\alpha_s\chi$ per unit rapidity.

The normalization $Z_{n,\nu}$ of the Born exchange of the BLV state in Eq.~\eqref{eq:blvborn} determines how the dipole couples near this maximum. Equation~\eqref{eq:Z} expresses it as the sum over even $n'$ of the integral of $P_{n,\nu}$ over $\nu'$. Both $\chi$ and the denominator $\mathcal D_1$ of Eq.~\eqref{eq:Ddef} vanish at $(1,0)$ when evaluated at the labels of the full state. The digamma series resolves their ratio,
\begin{align}
Z_{n,\nu}&=-\frac{3\pi}{2}\,\frac{\chi(n,\nu)}{\mathcal D_1(n,\nu)},\nonumber\\
\chi(1,\nu)&=-2\nu^2\sum_{k=1}^{\infty}\frac{1}{k(k^2+\nu^2)},\nonumber\\
2\pi Z_{\pm1,\nu}&=\frac{6\pi^2}{1+\nu^2}\sum_{k=1}^{\infty}\frac{1}{k(k^2+\nu^2)}\nonumber\\
&=6\pi^2\big[\zeta(3)-(\zeta(3)+\zeta(5))\nu^2+O(\nu^4)\big].
\label{eq:Zone}
\end{align}
In cubic exchange the conformal partial wave $\Psi_{\pm1,\nu}$ of Eq.~\eqref{eq:Phidef} carries the labels of the full state. Its LO coefficient in Eq.~\eqref{eq:blvsuper} equals $(\nu^2+1/4)/\pi^4$ times $2\pi Z_{\pm1,\nu}$. In our normalization this coefficient is analytic in the strip $|\mathrm{Im}\,\nu|<1$ and positive for real $\nu$. In particular it is finite and nonzero at $\nu=0$, where the expression in Ref.~\cite{KovchegovSzymanowskiWallon:2004} has the form $0/0$.

Besides the BLV state, the LO $C$-odd three-gluon sector contains a second state with intercept $1-\nOddStateCoeff\,\alpha_s N_c/\pi$~\cite{JanikWosiek:1999}. Two gluons necessarily coincide at one end when three gluons attach to one dipole. The wave function of this state vanishes at coincidence, hence its dipole coupling is zero~\cite{HattaIancuItakuraMcLerran:2005}.

\section{The quark term and its color algebra}
\label{sm:color}
We obtain the color coefficient in Eq.~\eqref{eq:quark} by expanding the quark term of the NLO dipole equation~\cite{BalitskyChirilli:2008}. First, the phase $\theta$ in Eq.~\eqref{eq:Kq} enters the coordinate-space kernel through the identity
\begin{align}
&|x-z'|^2|y-z|^2+|y-z'|^2|x-z|^2-|x-y|^2|z-z'|^2\nonumber\\
&\quad=2\,|x-z|\,|y-z'|\,|x-z'|\,|y-z|\cos\theta.
\label{eq:Kqid}
\end{align}
The kernel multiplies $c_q S_q$, with the second trace at coincident emission points subtracted. We choose fundamental generators with $\mathrm{tr}(t^at^b)=\delta^{ab}/2$. Expanding Eq.~\eqref{eq:Sqquad} gives the cubic imaginary part denoted by the arrow,
\begin{align}
S_q(z,z')&\to-\frac{6}{4N_c}\,T(D_{xy},D_{zz'},D_{zz'})\nonumber\\
&\quad+\frac{N_c}{4}\big[C(D_{xz'})-C(D_{xz})\big]-\frac{1}{4N_c}\,C(D_{zz'})\nonumber\\
&\quad-\frac{6}{4N_c}\,T(D_{xy},D_{xy},D_{zz'}).
\label{eq:Bquark}
\end{align}
The difference of the two planar dipoles is odd under $z\leftrightarrow z'$, as are the terms cubic and linear in $D_{zz'}$. Their integrals vanish against $K_q$. In the even term the generators alternate with the two field insertions around the quark loop. Before expanding the outer trace, the corresponding part is
\begin{align}
S_q(z,z')\big|_{\rm even}&=-\tfrac12\,\mathrm{Tr}\{t^aU_xt^bU_y^\dagger\}\nonumber\\
&\quad\times\mathrm{tr}\big(t^aD_{zz'}t^bD_{zz'}\big).
\label{eq:Beven}
\end{align}
The constant outer trace $\delta^{ab}/2$ selects a color-singlet pair. The symmetric term linear in the field, $i d^{abc}D_{xy}^c/4$, selects a symmetric adjoint pair. The contracted inner trace is the color factor of the non-planar loop. For the color-singlet and symmetric adjoint pairs, respectively, we find
\begin{align}
\delta^{ab}\,\mathrm{tr}(t^at^ct^bt^d)&=-\frac{\delta^{cd}}{4N_c},\nonumber\\
d^{abe}\,\mathrm{tr}(t^at^ct^bt^d)&=-\frac{d^{cde}}{2N_c}.
\label{eq:Bcross}
\end{align}
The color factor of the non-planar loop on a symmetric adjoint pair, together with the three choices of pair, accounts for the coefficient multiplying the cubic form in Eq.~\eqref{eq:quark}. To fix the normalization of $K_q$ against the color-singlet BFKL eigenvalue, we use the value of that color factor on a color-singlet pair in the evolution of the quadratic dipole,
\begin{align}
\frac{d}{dY}\,\big(D_{xy}^aD_{xy}^a\big)\Big|_q=-\frac{c_q}{4N_c}&\int d^2z\,d^2z'\,K_q(x,y;z,z')\nonumber\\
&\times D_{zz'}^aD_{zz'}^a.
\label{eq:Bsinglet}
\end{align}
Let $\lambda_q(n',\nu')$ be the eigenvalue obtained by integrating $K_q$ over the $q\bar q$ positions against the gluon-pair conformal eigenfunction $E^{n',\nu'}$. At quadratic order Eq.~\eqref{eq:Bsinglet} exhausts the terms of order $n_f\alpha_s^2/N_c$ in the linearized NLO dipole equation of Ref.~\cite{BalitskyChirilli:2008}. The eigenvalue of this term on a color-singlet pair is $-c_q\lambda_q/(4N_c)$. In the NLO BFKL eigenvalue of Kotikov and Lipatov~\cite{KotikovLipatov:2000} the corresponding non-planar quark term is $-(n_f\alpha_s^2/4N_c)\,w(n',1/2+i\nu')$. Matching the singlet values fixes $\lambda_q=4\pi^4 w$ for even $n'$, and integration over the reference point yields Eq.~\eqref{eq:Kqw}.

The symmetric and antisymmetric adjoint pair kernels of Ref.~\cite{BartelsFadinLipatovVacca:2013} differ by the Abelian part. For the non-planar quark loop its eigenvalue is twice the singlet value by Eq.~\eqref{eq:Bcross}. The connected three-to-three kernel at Born order is gluonic, so the quark term in that formulation is in the pair kernels. In their normalization the color coefficient $b_f$ of the Abelian part for fundamental Dirac quarks follows from the same traces,
\begin{align}
b_f&=\frac{n_fN_c}{2(N_c^2-4)(N_c^2-1)}\,d^{eab}d^{ecd}\nonumber\\
&\quad\times\big[\mathrm{tr}(t^at^dt^bt^c)+\mathrm{tr}(t^bt^dt^at^c)\big]\nonumber\\
&=-\frac{n_f}{2N_c}.
\label{eq:Bbf}
\end{align}
For adjoint matter this coefficient vanishes on a symmetric adjoint pair~\cite{BartelsFadinLipatovVacca:2013}, and the cubic contribution is a superposition of dipoles.

We next reduce the SU(3) quark term in Eq.~(8.10) of Ref.~\cite{BalitskyGrabovsky:2015} to dipole differences. Denote the three-quark Wilson loop by $\mathcal U$ and its part odd under conjugation of the Wilson lines by $\mathcal U^-$. Contracting the fundamental color indices with the two Levi-Civita tensors gives a dipole when two positions coincide,
\begin{align}
\mathcal U_{x_1x_2x_3}&=\varepsilon^{i'j'k'}\varepsilon_{ijk}\,(U_{x_1})_{i'}^{\ i}\,(U_{x_2})_{j'}^{\ j}\,(U_{x_3})_{k'}^{\ k},\nonumber\\
\mathcal U^-_{x_1x_2x_3}&=\mathcal U_{x_1x_2x_3}-\mathcal U_{x_1x_2x_3}\big|_{U\to U^\dagger},\nonumber\\
\mathcal U_{xyy}&=2\,\mathrm{Tr}\{U_xU_y^\dagger\}.
\label{eq:Uloop}
\end{align}
We identify the rapidity in the SU(3) equation with $Y$ because its LO term at cubic order coincides with the LO dipole term $\bar\alpha_s K_0/(2\pi)$ at $N_c=3$. In the three-gluon approximation its quark term for the dipole loop involves both coincident and distinct emission points,
\begin{align}
\frac{d}{dY}\,\mathcal U^-_{xyy}\Big|_q&=\frac{\alpha_s^2n_f}{48\pi^4}\int d^2z\,d^2z'\,K_q(x,y;z,z')\nonumber\\
&\quad\times\big[2\mathcal U^-_{xzz'}-\mathcal U^-_{xzz}-\mathcal U^-_{xz'z'}-(x\to y)\big].
\label{eq:Uquark}
\end{align}
Here we have expressed the kernel of Ref.~\cite{BalitskyGrabovsky:2015} in terms of $K_q$. The relative sign and factor follow by writing the transverse distances through $\tau$ and $\theta$,
\begin{widetext}
\begin{align}
&\frac{1}{|z-z'|^4}\bigg[\frac{|x-z'|^2|y-z|^2+|x-z|^2|y-z'|^2-|x-y|^2|z-z'|^2}{2\big(|x-z'|^2|y-z|^2-|x-z|^2|y-z'|^2\big)}
\ln\frac{|x-z'|^2|y-z|^2}{|x-z|^2|y-z'|^2}-1\bigg]\nonumber\\
&\qquad=\frac{1}{|z-z'|^4}\Big[\frac{\tau}{\sinh\tau}\cos\theta-1\Big]=-\frac14\,K_q(x,y;z,z').
\label{eq:Ukernel}
\end{align}
\end{widetext}
Expanding the loop to cubic order cancels the antisymmetric color tensor. The surviving symmetric tensor yields the trilinear form
\begin{align}
\mathcal U^-_{x_1x_2x_3}&=2i\,T\big(D_{x_1x_2}+D_{x_1x_3},\,D_{x_2x_1}+D_{x_2x_3},\nonumber\\
&\qquad\qquad D_{x_3x_1}+D_{x_3x_2}\big)+O(A^5),\nonumber\\
\mathcal U^-_{xyy}&=4i\,C(D_{xy})+O(A^5).
\label{eq:Ucubic}
\end{align}
Combining these forms at the emission points in the integrand reduces the difference of Wilson loops to
\begin{align}
&2\mathcal U^-_{xzz'}-\mathcal U^-_{xzz}-\mathcal U^-_{xz'z'}-(x\to y)\nonumber\\
&\qquad=-24i\,T(D_{xy},D_{zz'},D_{zz'})+O(A^5).
\label{eq:Ubracket}
\end{align}
The cubic parent carries the factor $4i$ in Eq.~\eqref{eq:Ucubic}. Upon division by this factor we recover Eq.~\eqref{eq:quark} at three colors, since $6c_q/(4N_c)=\alpha_s^2 n_f/(8\pi^4)$ for SU(3).

\section{Conservation of $L_g$ in the NLO dipole equation}
\label{sm:charge}
We can evaluate $L_g$ directly on the traces of the dipole equation. Let $N_t$ be the number of traces and $Q_{m,x}$ the sum of the signs of the Wilson lines at $x$ in trace $m$. A trace has zero linear term. Thus the three fields of a local cubic term must all come from the same trace, and the remaining traces contribute their constant parts. We obtain
\begin{equation}
L_g\Big[\prod_{m=1}^{N_t}\mathrm{Tr}_m\Big]=N_c^{\,N_t-1}\sum_{m=1}^{N_t}\sum_x g(x)\,Q_{m,x}^{3},
\label{eq:Btrace}
\end{equation}
which can be applied before integrating over the emissions. At cubic order each gluon structure of Ref.~\cite{BalitskyChirilli:2008} reduces to a superposition of dipoles. We write the three structures in terms of traces, reduce them to dipole forms at cubic order and evaluate $L_g$ on those forms,
\begin{align}
S_1&=\mathrm{Tr}\{U_xU_z^\dagger\}\mathrm{Tr}\{U_zU_y^\dagger\}-N_c\,\mathrm{Tr}\{U_xU_y^\dagger\}\nonumber\\
&\to N_c\big[C(D_{xz})+C(D_{zy})-C(D_{xy})\big],\nonumber\\
S_2&=\mathrm{Tr}\{U_xU_z^\dagger\}\mathrm{Tr}\{U_zU_{z'}^\dagger\}\mathrm{Tr}\{U_{z'}U_y^\dagger\}\nonumber\\
&\quad-\mathrm{Tr}\{U_xU_z^\dagger U_{z'}U_y^\dagger U_zU_{z'}^\dagger\}-(z'\to z)\nonumber\\
&\to N_c^2\big[C(D_{zz'})+C(D_{z'y})-C(D_{zy})\big],\nonumber\\
S_3&=\mathrm{Tr}\{U_xU_z^\dagger\}\mathrm{Tr}\{U_zU_{z'}^\dagger\}\mathrm{Tr}\{U_{z'}U_y^\dagger\}\nonumber\\
&\to N_c^2\big[C(D_{xz})+C(D_{zz'})+C(D_{z'y})\big],\nonumber\\
&L_g[S_1]=L_g[S_2]=0,\quad L_g[S_3]=N_c^2\big[g(x)-g(y)\big].
\label{eq:Bgluon}
\end{align}
Every term with a single integration point multiplies $S_1$. Its functional vanishes at each emission point because the daughter coefficients cancel there and the virtual term cancels the contributions at the parent ends, as in Eq.~\eqref{eq:telescope}. The first logarithm in the single-integral kernel of the dipole equation contains the normalization point $\hat\mu$ of Ref.~\cite{BalitskyChirilli:2008}. We denote the coefficient in that reference by $\beta_0$ (Sec.~\ref{sec:shift}) and write the kernel in the form
\begin{align}
K_1(x,y;z)={}&\frac{|x-y|^2}{|x-z|^2|y-z|^2}\bigg\{1+\frac{\alpha_s}{4\pi}\bigg[\beta_0\ln\big(|x-y|^2\hat\mu^2\big)\nonumber\\
&-\beta_0\,\frac{|x-z|^2-|y-z|^2}{|x-y|^2}\,\ln\frac{|x-z|^2}{|y-z|^2}\nonumber\\
&+\Big(\frac{67}{9}-\frac{\pi^2}{3}\Big)N_c-\frac{10}{9}\,n_f\nonumber\\
&-2N_c\ln\frac{|x-z|^2}{|x-y|^2}\,\ln\frac{|y-z|^2}{|x-y|^2}\bigg]\bigg\}.
\label{eq:K1}
\end{align}
The $1$ in braces gives the LO kernel of Eq.~\eqref{eq:K0}.

The six-line trace in $S_2$ has the same difference $D_{xy}$ before and after setting $z'=z$, so the coincidence subtraction removes its cubic part. The remaining dipoles in $S_2$ have zero functional for each $z,z'$. The kernel of $S_2$ is even under interchange of $z$ and $z'$. For distinct points we write it in the variables $\tau,\theta$ of Eq.~\eqref{eq:Kq},
\begin{align}
K_2(x,y;z,z')={}&\frac{4}{|z-z'|^4}\,\big(\tau\coth\tau\,\cos2\theta\nonumber\\
&-2\tau\sinh\tau\,\cos\theta-1\big).
\label{eq:K2}
\end{align}
For $S_3$ the functional depends on the parent ends alone. We write the kernel of the $S_3$ term in the same variables,
\begin{align}
\frac{d}{dY}\,\mathrm{Tr}\{U_xU_y^\dagger\}\Big|_{S_3}&=\frac{\alpha_s^2}{16\pi^4}\int d^2z\,d^2z'\nonumber\\
&\quad\times K_3(x,y;z,z')\,S_3(z,z'),\nonumber\\
K_3(x,y;z,z')&=\frac{8}{|z-z'|^4}\,\tau\cos\theta\,\big(\cosh\tau-\cos\theta\big).
\label{eq:K3}
\end{align}
The kernel $K_3$ is odd under interchange of $z$ and $z'$. Its integral against the constant functional of $S_3$ is the one conditionally convergent integral here.
The derivation of the dipole equation uses the same longitudinal cutoff for the two gluon orderings, and the cross terms of the two-gluon emission give the antisymmetrized combination on the right below~\cite{BalitskyChirilli:2008}. The symmetric prescription follows from this antisymmetrized form. To obtain the product on the left we relabel one of its two integrals, which is valid only for a regularization symmetric under interchange of the emission points. The two forms then obey
\begin{align}
\int d^2z\,d^2z'\,K_3\,S_3&=\tfrac12\int d^2z\,d^2z'\,K_3(x,y;z,z')\nonumber\\
&\quad\times\big[S_3(z,z')-S_3(z',z)\big],\nonumber\\
S_3(z,z')-S_3(z',z)&\to N_c^2\big[C(D_{xz})-C(D_{xz'})\nonumber\\
&\quad+2C(D_{zz'})+C(D_{z'y})-C(D_{zy})\big].
\label{eq:S3sym}
\end{align}
The difference of the two orderings has vanishing functional for each pair of emission points. We can also integrate over $z'$ first in the symmetric prescription. The integral vanishes away from the parent ends, but as a distribution in $z$ it contains the delta-function contact terms found in Ref.~\cite{BalitskyChirilli:2009}. Normalizing each delta function to unit integral against $d^2z$, we obtain
\begin{equation}
\int d^2z'\,K_3(x,y;z,z')=4\pi^2\zeta(3)\,\big[\delta^2(z-y)-\delta^2(z-x)\big].
\label{eq:S3contact}
\end{equation}
Both contacts multiply the same functional of $S_3$. Their opposite signs therefore give the cancellation
\begin{align}
&\int d^2z\,d^2z'\,K_3(x,y;z,z')\,L_g[S_3]\nonumber\\
&\quad=N_c^2\big[g(x)-g(y)\big]\,4\pi^2\zeta(3)\nonumber\\
&\qquad\times\int d^2z\,\big[\delta^2(z-y)-\delta^2(z-x)\big]=0.
\label{eq:S3cancel}
\end{align}
For the quark term we denote the fundamental Casimir by $C_F=(N_c^2-1)/(2N_c)$. Applying Eq.~\eqref{eq:Lg} to the cubic expansion~\eqref{eq:Bquark} yields
\begin{align}
L_g\big[S_q(z,z')\big]&=\Big(\frac{N_c}{4}-\frac{1}{4N_c}\Big)\big[g(z)-g(z')\big]\nonumber\\
&=\frac{C_F}{2}\big[g(z)-g(z')\big].
\label{eq:Bcharge}
\end{align}
For distinct ends no point belongs to all three arguments of a trilinear term. At each point at least one argument has zero net sign, so every such term vanishes under $L_g$. The difference that remains is odd in $z,z'$, so its absolutely convergent integral against $K_q$ is zero.

For an adjoint line the fundamental line and its conjugate at the emission point give zero net sign. The subtraction at the parent ends completes $S_1$ for one integration point. The two-emission terms for adjoint scalars and fermions have the subtracted structure $S_2$~\cite{BalitskyChirilli:2009}, so these terms also conserve $L_g$.

\section{First-order perturbation theory and the left eigenvector}
\label{sm:firstorder}
We derive the first-order ratio by finding a right eigenvector at the prescribed labels and a left eigenvector valid on the correction to the state. The non-planar quark term produces the non-dipole state of Eq.~\eqref{eq:quark}. The corresponding state correction $W^{(1)}_{n,\nu}$ therefore has a non-dipole component. By Eq.~\eqref{eq:lefteig} the overlap $\Pi_{n,\nu}$ is a left eigenvector of $H_0$ on cubic forms built from dipole differences and their emission integrals, whether they are superpositions of dipoles or non-dipole forms. Together the two eigenvectors give Eq.~\eqref{eq:firstorder} under the isolation condition of Sec.~\ref{sec:shift}.

Consider first the LO Hamiltonian acting on the integral defining $\mathcal B_{n,\nu}$ in Eq.~\eqref{eq:blvform}. The real terms and their virtual subtraction diverge separately at the ends of the integration dipole. We remove equal disks at both ends in each term, together with integration dipoles shorter than the common radius. The invariant dipole measure times the real-emission kernel has a density symmetric in the two ends and the emission point. We may therefore relabel the real terms, while the virtual subtraction is diagonal. The kernel now acts on the conjugate eigenfunction, so
\begin{align}
H_0\,\mathcal B_{n,\nu}(x_0)&=\frac{\bar\alpha_s}{2\pi}\int\frac{d^2x_1\,d^2x_2}{|x_1-x_2|^4}\,\big(K_0E^{n,\nu\,*}\big)(x_1,x_2;x_0)\nonumber\\
&\quad\times C(D_{x_1x_2})=\bar\alpha_s\chi(n,\nu)\,\mathcal B_{n,\nu}(x_0).
\label{eq:blveigen}
\end{align}
Replacing the dipole form in the defining integral by its exchange $l^3$ determines the Born exchange of the BLV state. Its coefficient follows from the amplitude for the three-gluon state of a dipole~\cite{KovchegovSzymanowskiWallon:2004},
\begin{align}
F_{\mathcal B_{n,\nu}(x_0)}(u,v)&=\int\frac{d^2x_1\,d^2x_2}{|x_1-x_2|^4}\nonumber\\
&\quad\times E^{n,\nu}(x_1,x_2;x_0)^{*}\,l(x_1,x_2;u,v)^3\nonumber\\
&=2\pi Z_{n,\nu}\,E^{n,\nu}(u,v;x_0)^{*},\nonumber\\
Z_{n,\nu}&=-\frac{3\pi}{2}\frac{\chi(n,\nu)}{\mathcal D_1(n,\nu)}.
\label{eq:blvborn}
\end{align}
Conversely, we can resolve a dipole with specified ends into these states. The completeness relation in the invariant dipole measure is
\begin{align}
&\delta^2(x-x_1)\,\delta^2(y-x_2)\nonumber\\
&\quad=\sum_{n=-\infty}^{\infty}\int_{-\infty}^{\infty}d\nu\,\frac{\nu^2+n^2/4}{\pi^4\,|x_1-x_2|^4}\nonumber\\
&\qquad\times\int d^2x_0\,E^{n,\nu}(x,y;x_0)\,E^{n,\nu}(x_1,x_2;x_0)^{*}.
\label{eq:completeness}
\end{align}
Exchanging the two ends changes the sign of the three-gluon state of a dipole and multiplies the conformal eigenfunction by $(-1)^n$, so only odd spins survive the integration. We obtain the superposition for the form and for its Born exchange, where $\Psi_{n,\nu}$ is the conformal partial wave~\eqref{eq:Phidef} at the full-state labels,
\begin{align}
C(D_{xy})&=\sum_{n\ \mathrm{odd}}\int d\nu\,\frac{\nu^2+n^2/4}{\pi^4}\nonumber\\
&\quad\times\int d^2x_0\,E^{n,\nu}(x,y;x_0)\,\mathcal B_{n,\nu}(x_0),\nonumber\\
l(x,y;u,v)^3&=\sum_{n\ \mathrm{odd}}\int d\nu\,\frac{\nu^2+n^2/4}{\pi^4}\;2\pi Z_{n,\nu}\nonumber\\
&\quad\times\Psi_{n,\nu}(x,y;u,v).
\label{eq:blvsuper}
\end{align}
Combining opposite labels with the same Casimir eigenvalues gives $\mathcal C_{n,\nu}$ of Eq.~\eqref{eq:Cright}. We label another such sector by $(\tilde n,\tilde\nu)$. Since the target integration commutes with differentiation by the projectile Casimir operators, its overlap vanishes on forms with different Casimir eigenvalues. A conformally invariant $V$ commutes with M\"obius transformations of all gluon coordinates and is diagonal in those labels. The same selection therefore applies after insertion of $V$,
\begin{align}
\Pi_{n,\nu}\big[\mathcal C_{\tilde n,\tilde\nu}\big]&=\delta_{n\tilde n}\,\delta(\nu-\tilde\nu)\,\Pi_{n,\nu}\big[C(D_{xy})\big],\nonumber\\
\Pi_{n,\nu}\big[V\,\mathcal C_{\tilde n,\tilde\nu}\big]&=\delta_{n\tilde n}\,\delta(\nu-\tilde\nu)\,\Pi_{n,\nu}\big[V\,C(D_{xy})\big]\nonumber\\
&\qquad(n,\tilde n\geq1).
\label{eq:labelsel}
\end{align}
We apply the overlap $\Pi_{n,\tilde\nu}$ to Eq.~\eqref{eq:firstordereq}. The correction $W^{(1)}_{n,\nu}$ lies at the chosen labels because $H_0$ and the conformally invariant perturbation $V$ preserve them in the first-order equation. Label selection therefore makes the overlap of the correction vanish unless $\tilde\nu=\nu$. At equal labels the difference of LO eigenvalues left by Eq.~\eqref{eq:lefteigid} is zero, so the term with the state correction drops out. Both terms on the right have the delta function from Eq.~\eqref{eq:labelsel}, leaving
\begin{align}
0=\delta(\tilde\nu-\nu)\Big\{&\omega^{(1)}_V(n,\nu)\,\Pi_{n,\nu}\big[C(D_{xy})\big]\nonumber\\
&-\Pi_{n,\nu}\big[V\,C(D_{xy})\big]\Big\}.
\label{eq:firstorderproj}
\end{align}
Its coefficient must vanish. This fixes the eigenvalue correction and permits the state correction to be chosen with zero component on the BLV state. Equation~\eqref{eq:firstordereq} determines the component on each level whose eigenvalue differs from the BLV eigenvalue. The integral over the rapidity $Y'$ converges only if every level that $V$ mixes with the BLV state lies below the BLV eigenvalue. The evolution of the remaining components with $H_0$ then gives
\begin{equation}
W^{(1)}_{n,\nu}=\int_0^\infty dY'\;e^{Y'[H_0-\bar\alpha_s\chi(n,\nu)]}\,\big[V-\omega^{(1)}_V(n,\nu)\big]\,\mathcal C_{n,\nu}.
\label{eq:W1}
\end{equation}
At $(1,0)$ the other levels of zero quasimomentum found in the numerical search had eigenvalues below that of the BLV state (Sec.~\ref{sec:shift} and Appendix~\ref{sm:levels}). The components of $W^{(1)}_{n,\nu}$ on those levels consequently decay in the integrand with increasing $Y'$.

We now compute the overlap in Eq.~\eqref{eq:lefteig}. In Sec.~\ref{sec:shift} we combined the polarization identity~\eqref{eq:mixedpair}, Eq.~\eqref{eq:transfer} and the pair sum~\eqref{eq:pairsum} to obtain the Born exchange after $H_0$ acts on the cubic form $W$. At disk radius $\eta$ the symmetry of the regulated region and of the real-emission density lets us relabel the points in each real term to obtain
\begin{widetext}
\begin{align}
\Pi_{n,\nu}\big[H_0W\big]&=\frac{\bar\alpha_s}{2\pi}\int\frac{d^2u\,d^2v}{|u-v|^4}\,E^{n,\nu}\,K_0^{(u,v)}F_W\nonumber\\
&=\frac{\bar\alpha_s}{2\pi}\lim_{\eta\to0}\int_{|u-v|,\,|u-p|,\,|p-v|>\eta}\frac{d^2u\,d^2v\,d^2p}{|u-v|^2\,|u-p|^2\,|p-v|^2}\,E^{n,\nu}(u,v;x_0)\,\big[F_W(u,p)+F_W(p,v)-F_W(u,v)\big]\nonumber\\
&=\frac{\bar\alpha_s}{2\pi}\lim_{\eta\to0}\int_{|u-v|,\,|u-p|,\,|p-v|>\eta}\frac{d^2u\,d^2v\,d^2p}{|u-v|^2\,|u-p|^2\,|p-v|^2}\,F_W(u,v)\nonumber\\
&\qquad\times\big[E^{n,\nu}(u,p;x_0)+E^{n,\nu}(p,v;x_0)-E^{n,\nu}(u,v;x_0)\big]\nonumber\\
&=\bar\alpha_s\,\chi(n,\nu)\,\Pi_{n,\nu}[W].
\label{eq:lefteig}
\end{align}
\end{widetext}
For real $\nu$ the modulus of the eigenfunction vanishes linearly when the ends coincide.
Removing the common regulator in Eq.~\eqref{eq:lefteig} requires an absolutely convergent overlap.
The integral converges near coincident target ends if the Born exchange vanishes there with an exponent greater than one, even with logarithmic factors.
The Born exchanges of cubic forms built from dipole differences at fixed, distinct projectile points vanish cubically when the target ends meet away from those points.
The Born exchanges obtained by applying $H_0$ or the non-planar quark term to a fixed dipole $C(D_{xy})$ retain this cubic behavior up to powers of a logarithm from the emission integrals.
The singularities at the projectile ends are integrable powers of logarithms.
The Born exchange of $\mathcal C_{n,\nu}$ is proportional to the conformal partial wave in Eq.~\eqref{eq:blvsuper}.
The state correction $W^{(1)}_{n,\nu}$ has the same conformal labels, and its Born exchange is a combination of the conformal partial waves at those labels.
These conformal partial waves vanish only as the first power of the target dipole size up to logarithms, so we cannot assume absolute convergence at equal labels when applying Eq.~\eqref{eq:lefteigid} to the correction.
Its overlap is a distribution in the label, proportional to the delta function in Eq.~\eqref{eq:labelsel}.
A smooth superposition over the label suppresses the Born exchange at small target dipole size enough for the overlap to converge absolutely.
We can therefore remove the regulator before taking the limit to the chosen labels, which gives Eq.~\eqref{eq:firstorderproj} with the delta function common to both terms.
The full dipole enters the ratio for the non-planar quark term through $lR_q$ in the numerator and $l^3$ in the denominator.
Their cubic behavior up to logarithms makes both overlaps absolutely convergent without a regulator.
In the ratio~\eqref{eq:firstorder} the normalizations of the BLV state and conformal eigenfunction cancel, together with the Born color and coupling factor.

\section{Conformal partial wave expansion and overlaps}
\label{sm:pairs}
\subsection{The overlap at the branch point}
We first determine the coefficient of each conformal partial wave in $l^2$. The overlap $\Pi_{1,0}$ of an amplitude $F$ depending on $\rho$ alone reduces to an inner product of $l$ with $F$ on the cross-ratio plane whenever the overlap converges absolutely. This reduction also gives the norm $\langle l^2|l^2\rangle$. For the expansion we differentiate $l^2$ with the transverse Laplacians at $z$ and $z'$. At distinct gluon positions the cross term places a gluon at each target end. Terms with both gluons at the same end have no contribution there. We find
\begin{align}
\Delta_z\Delta_{z'}\,l(z,z';u,v)^2&=8\pi^2\big[\delta^2(z-u)\,\delta^2(z'-v)\nonumber\\
&\qquad+\delta^2(z-v)\,\delta^2(z'-u)\big].
\label{eq:l2waves}
\end{align}
The two Laplacians acting on a conformal partial wave multiply it by $16\mathcal D_1(n',\nu')/|z-z'|^4$ at distinct points. Expanding the delta functions by completeness~\cite{KovchegovSzymanowskiWallon:2004} then fixes the coefficient in Eq.~\eqref{eq:l2expand}.

For the overlap it is convenient to take the reference point to infinity. Removing the power dependence of the eigenfunction there leaves the integral
\begin{align}
&\lim_{x_0\to\infty}x_0^{2h}\bar x_0^{2\bar h}\,\Pi_{n,\nu}[F]\nonumber\\
&\quad=\int\frac{d^2u\,d^2v}{|u-v|^4}\,(u-v)^h(\bar u-\bar v)^{\bar h}\,F(x,y;u,v).
\label{eq:frame}
\end{align}
We put the projectile ends at 0 and 1. At the branch point the conformal weights are $h=1$ and $\bar h=0$. We take amplitudes $F$ that depend on the points only through $\rho$ and whose overlaps converge absolutely, such as $lR_q$ and $l^3$ in Sec.~\ref{sec:shift}. At fixed $u$ we can then replace $v$ by the cross ratio and take $F$ outside the remaining $u$ integral,
\begin{align}
v&=\frac{u(1-\rho)}{1-\rho u},\nonumber\\
\frac{d^2u\,d^2v}{|u-v|^4}&=\frac{d^2u\,d^2\rho}{|u|^2|1-u|^2|\rho|^4},\nonumber\\
\Pi_{1,0}[F]&=-E^{1,0}(x,y;x_0)\int\frac{d^2\rho}{|\rho|^4}\,F\nonumber\\
&\quad\times\int d^2u\,\frac{\rho}{\bar u(1-\bar u)(1-\rho u)}.
\label{eq:Pi0u}
\end{align}
The remaining integral over $u$ evaluates to $-2\pi l$, where $l=\ln|1-\rho|$ is the exchange~\eqref{eq:l} in these coordinates. For general real labels we use the same change of variables for amplitudes depending on $\rho$ alone whose overlaps converge absolutely. This includes $l\Psi_{n',\nu'}$, $l^3$, $lR_q$ and the factor $|z-z'|^4K_q$ of the quark kernel. Keeping the conformal weights explicit, the target integral becomes
\begin{align}
&\Pi_{n,\nu}[F](x,y;x_0)=E^{n,\nu}(x,y;x_0)\int\frac{d^2\rho}{|\rho|^4}\,F\;(-1)^n\rho^{h}\bar\rho^{\bar h}\nonumber\\
&\quad\times\int d^2u\,u^{h-1}\bar u^{\bar h-1}(1-u)^{h-1}(1-\bar u)^{\bar h-1}\nonumber\\
&\quad\times(1-\rho u)^{-h}(1-\bar\rho\bar u)^{-\bar h}.
\label{eq:reduce}
\end{align}
Rescaling the remaining target coordinate turns it into the reference point of the conformal partial wave at $(n,\nu)$. That function depends only on the cross ratio, so we can evaluate it with dipole ends $(1,\infty)$ and $(0,\rho)$ and obtain
\begin{align}
&(-1)^n\rho^{h}\bar\rho^{\bar h}\int d^2u\,u^{h-1}\bar u^{\bar h-1}(1-u)^{h-1}(1-\bar u)^{\bar h-1}\nonumber\\
&\quad\times(1-\rho u)^{-h}(1-\bar\rho\bar u)^{-\bar h}\nonumber\\
&=\Psi_{n,\nu}(1,\infty;0,\rho)\equiv\Psi_{n,\nu}(\rho),\nonumber\\
&\Pi_{n,\nu}[F]=\langle\Psi_{n,\nu}|F\rangle\,E^{n,\nu}(x,y;x_0).
\label{eq:Psirho}
\end{align}
At $(1,0)$ the integral gives $\Psi_{1,0}=2\pi l$, reducing the general overlap to Eq.~\eqref{eq:Pi0rho}. We use $r=|1-\rho|$ for the radial integration of the norm, so that $l=\ln r$. Performing the angular integral and inverting the radius for $r>1$ gives
\begin{align}
\langle l^2|l^2\rangle&=4\pi\int_0^1dr\,\frac{r(1+r^2)\ln^4r}{(1-r^2)^3}\nonumber\\
&=3\pi\zeta(3).
\label{eq:l2norm}
\end{align}
The factor $2\pi$ from the target integral converts this norm to the dipole coefficient in Eq.~\eqref{eq:Pi0dipole}.
\subsection{The overlap of one conformal partial wave}
The overlap at general real labels also has a closed form. We start by interchanging the target ends, which maps $\rho$ to $\rho'=\rho/(\rho-1)$. The conformal partial wave has the parity of its spin under this interchange, while $l$ changes sign and the measure is invariant,
\begin{align}
\Psi_{n,\nu}\Big(\frac{\rho}{\rho-1}\Big)&=(-1)^n\Psi_{n,\nu}(\rho),\qquad
l\Big(\frac{\rho}{\rho-1}\Big)=-l(\rho),\nonumber\\
\frac{d^2\rho'}{|\rho'|^4}&=\frac{d^2\rho}{|\rho|^4}.
\label{eq:exchange}
\end{align}
Thus the overlap of $l\Psi_{n',\nu'}$ with $\Psi_{n,\nu}$ vanishes for even $n+n'$. At odd $n+n'$ we evaluate it using the holomorphic and antiholomorphic Casimir operators. Write $h',\bar h'$ for the weights of Eq.~\eqref{eq:Enu} at the pair labels. The holomorphic eigenvalue and the difference of the two holomorphic eigenvalues follow from
\begin{align}
(x-y)^2\partial_x\partial_y\,E^{n,\nu}(x,y;x_0)&=h(1-h)\,E^{n,\nu}(x,y;x_0),\nonumber\\
h(1-h)-h'(1-h')&=-(h-h')(h+h'-1).
\label{eq:CasEig}
\end{align}
To differentiate under the reference-point integral we take the $y$ derivative first near $x_0=x$ and the $x$ derivative first near $x_0=y$. The two Casimir equations for the conformal partial wave then take the cross-ratio form
\begin{align}
-\rho^2\partial_\rho\big[(1-\rho)\,\partial_\rho\Psi_{n,\nu}\big]&=h(1-h)\,\Psi_{n,\nu},\nonumber\\
-\bar\rho^{\,2}\partial_{\bar\rho}\big[(1-\bar\rho)\,\partial_{\bar\rho}\Psi_{n,\nu}\big]&=\bar h(1-\bar h)\,\Psi_{n,\nu}.
\label{eq:Casrho}
\end{align}
We integrate both equations on the same punctured plane, inside a large circle and outside small disks at 0 and 1. The boundary contributions require the behavior of $\Psi$ at the three punctures. The logarithm at infinity comes from the annulus $1\ll|x_0|\ll|\rho|$ in the defining integral. The interchange~\eqref{eq:exchange} maps the neighborhood of 1 to infinity and fixes the logarithmic coefficient there. The local solutions of the Casimir equations at zero bound the behavior of $\Psi$ as $\rho\to0$,
\begin{align}
\Psi_{n,\nu}&=\pi\ln|\rho|^2+O(1)\qquad(\rho\to\infty),\nonumber\\
\Psi_{n,\nu}&=-(-1)^n\pi\ln|1-\rho|^2+O(1)\qquad(\rho\to1),\nonumber\\
\Psi_{n,\nu}&=O\big(|\rho|\,(1+|\ln|\rho||)\big)\qquad(\rho\to0).
\label{eq:punct}
\end{align}
The logarithm $l$ obeys $(1-\rho)\partial_\rho l=-1/2$ away from the punctures. Differentiating the product of $l$ and the conformal partial wave therefore leaves one term in addition to the Casimir acting on $\Psi$,
\begin{align}
&-\rho^2\partial_\rho\big[(1-\rho)\,\partial_\rho(l\,\Psi)\big]\nonumber\\
&\quad=l\,\Big(-\rho^2\partial_\rho\big[(1-\rho)\,\partial_\rho\Psi\big]\Big)+\rho^2\,\partial_\rho\Psi.
\label{eq:ibp}
\end{align}
We integrate this equation against $\Psi_{n,\nu}$ in the cross-ratio measure. The holomorphic operator is symmetric with respect to this measure and Eq.~\eqref{eq:punct} makes the first boundary terms vanish. We thus obtain the difference of holomorphic eigenvalues times the overlap,
\begin{align}
&\big[h(1-h)-h'(1-h')\big]\,\langle\Psi_{n,\nu}|\,l\,\Psi_{n',\nu'}\rangle\nonumber\\
&\quad=\int d^2\rho\;\frac{\Psi_{n,\nu}\,\partial_\rho\Psi_{n',\nu'}}{\bar\rho^{\,2}}.
\label{eq:ibp1}
\end{align}
The antiholomorphic operator commutes with the remaining derivative $\partial_\rho$. Integrating by parts a second time converts the product of eigenvalue differences times the overlap to a boundary integral. The contours run clockwise around the two holes and counterclockwise around the outer circle,
\begin{align}
&\big[h(1-h)-h'(1-h')\big]\big[\bar h(1-\bar h)-\bar h'(1-\bar h')\big]\nonumber\\
&\quad\times\langle\Psi_{n,\nu}|\,l\,\Psi_{n',\nu'}\rangle\nonumber\\
&=-\frac{1}{2i}\oint d\rho\,(1-\bar\rho)\Big[(\partial_{\bar\rho}\Psi_{n,\nu})\,\partial_\rho\Psi_{n',\nu'}\nonumber\\
&\qquad-\Psi_{n,\nu}\,\partial_\rho\partial_{\bar\rho}\Psi_{n',\nu'}\Big].
\label{eq:ibp2}
\end{align}
The inner circle at zero gives a vanishing contribution as its radius shrinks. At 1 and at infinity the logarithmic coefficients fix the two integrals, and the relative orientation sets their signs. Removing the common regulator gives
\begin{align}
&-\frac{1}{2i}\oint d\rho\,(1-\bar\rho)\Big[\dots\Big]\nonumber\\
&\quad=\pi\Big[\pi\cdot\pi-\big(-(-1)^n\pi\big)\big(-(-1)^{n'}\pi\big)\Big]\nonumber\\
&\quad=\pi^3\big[1-(-1)^{n+n'}\big].
\label{eq:flux}
\end{align}
At real labels the product of the Casimir eigenvalue differences gives $\mathcal D_-\mathcal D_+$ of Eq.~\eqref{eq:Dmp}, the denominator in Eq.~\eqref{eq:single},
\begin{align}
&\big[h(1-h)-h'(1-h')\big]\big[\bar h(1-\bar h)-\bar h'(1-\bar h')\big]\nonumber\\
&\quad=\mathcal D_-\mathcal D_+.
\label{eq:CasDiff}
\end{align}
For odd $n+n'$ both factors are positive, so their product can be divided out. For equal parities Eq.~\eqref{eq:exchange} fixes the overlap even when the eigenvalues coincide. Combining the two cases gives
\begin{align}
\langle\Psi_{n,\nu}|\,l\,\Psi_{n',\nu'}\rangle&=\frac{2\pi^3}{\mathcal D_-\mathcal D_+}\qquad(n+n'\ \text{odd}),\nonumber\\
\langle\Psi_{n,\nu}|\,l\,\Psi_{n',\nu'}\rangle&=0\qquad(n+n'\ \text{even}).
\label{eq:singlerho}
\end{align}
At real labels the modulus of the integrand in the reference-point integral for $\Psi$ is independent of the labels. Integrating that modulus bounds each conformal partial wave, with the logarithmic behavior at the punctures in Eq.~\eqref{eq:punct}. The integrals for the overlap converge absolutely throughout the real labels, including $\nu=0$ and $\nu'=0$. Equation~\eqref{eq:Psirho} therefore yields the overlap of $l\Psi_{n',\nu'}$ with the BLV state in Eq.~\eqref{eq:single} even at the branch point. There $\Psi_{1,0}=2\pi l$ and $\mathcal D_-\mathcal D_+=\mathcal D_1(n',\nu')$. Thus at $(1,0)$ the overlap of $\Psi_{n',\nu'}$ with $l^2$ is $\pi^2/\mathcal D_1(n',\nu')$ for even $n'$, the factor in Eq.~\eqref{eq:P10}. It vanishes for odd $n'$, including at $n'=\pm1$ and $\nu'=0$.

Each conformal partial wave in the Born expansion~\eqref{eq:l2expand} acquires its own factor $\lambda_q$ under the quark kernel. Multiplying by the overlap just derived distinguishes the two target integrals by this factor alone,
\begin{align}
\Pi_{n,\nu}\big[l\,R_q\big]&=2\pi\,E^{n,\nu}(x,y;x_0)\nonumber\\
&\quad\times\sum_{n'\,\mathrm{even}}\int d\nu'\,P_{n,\nu}(n',\nu')\,\lambda_q(n',\nu'),\nonumber\\
\Pi_{n,\nu}\big[l^3\big]&=2\pi\,E^{n,\nu}(x,y;x_0)\sum_{n'\,\mathrm{even}}\int d\nu'\,P_{n,\nu}(n',\nu').
\label{eq:projKq}
\end{align}
Their ratio yields Eq.~\eqref{eq:avg}. The normalization in the amplitude for the three-gluon state of a dipole follows from its expansion in Ref.~\cite{KovchegovSzymanowskiWallon:2004}, where the sign used for one-gluon exchange is opposite to ours. Also, additivity of the exchange logarithm between the daughters relates the LO kernel on $l^3$ to that on $l^2$. In our conventions the relations are
\begin{align}
Z_{n,\nu}\equiv\sum_{n'\,\mathrm{even}}\int d\nu'\,P_{n,\nu}(n',\nu')&=-\frac{3\pi}{2}\,\frac{\chi(n,\nu)}{\mathcal{D}_1(n,\nu)},\nonumber\\
K_0\big[l^3\big]&=\tfrac32\,l\,K_0\big[l^2\big].
\label{eq:Z}
\end{align}
Applying $\Pi_{n,\nu}$ to the kernel relation and using the expansion~\eqref{eq:projKq} proves the LO identity~\eqref{eq:LOid}. To obtain $\lambda_q$ itself we use the reduction~\eqref{eq:Psirho} for the quark kernel, integrating over the $q\bar q$ positions in place of the target ends. The integral over one position at fixed cross ratio produces the conformal partial wave. The kernel has $\tau=-l$ and $\theta=-\arg(1-\rho)$, so its eigenvalue reduces to
\begin{equation}
\lambda_q(n',\nu')=\Big\langle 4\Big[1-\frac{l}{\sinh l}\,\cos\arg(1-\rho)\Big]\,\Big|\,\Psi_{n',\nu'}\Big\rangle.
\label{eq:Blambda}
\end{equation}

\section{Closed forms and the dependence on $\nu$}
\label{sm:closed}
We obtain the non-planar quark contribution to the $n=1$ eigenvalue used in Sec.~\ref{sec:location} from the average $\langle w\rangle_P$ at the branch point and $\langle w\rangle_P(\nu)$ along this line. The numerator of $\langle w\rangle_P$ at the branch point receives contributions from $n'=0$ and $n'=\pm2$ alone because $w$ vanishes at every other even spin. Denote $P_{1,0}$ by $P$ in the integrals at $(1,0)$. Evaluating the distribution and the function to be averaged at the contributing spins yields
\begin{align}
P(0,\nu')&=\frac{\nu'^2}{(\nu'^2+\frac14)^4},\nonumber\\
P(\pm2,\nu')&=\frac{\nu'^2+1}{(\nu'^2+\frac14)^2(\nu'^2+\frac94)^2},\nonumber\\
w\big(0,\tfrac12+i\nu'\big)&=\frac{\sinh(\pi\nu')\,(\frac{11}{4}+3\nu'^2)}{8\nu'\cosh^2(\pi\nu')\,(1+\nu'^2)},\nonumber\\
w\big(\pm2,\tfrac12+i\nu'\big)&=-\frac{\sinh(\pi\nu')\,(\frac14+\nu'^2)}{16\nu'\cosh^2(\pi\nu')\,(1+\nu'^2)}.
\label{eq:Pw10}
\end{align}
The integrals for $n'=\pm2$ coincide. We decompose the rational factors in partial fractions and sum the residues for $n'=0$ and $n'=2$. Writing $G$ for Catalan's constant, we obtain
\begin{align}
I_0&\equiv\int_{-\infty}^{\infty}d\nu'\,P(0,\nu')\,w\big(0,\tfrac12+i\nu'\big)\nonumber\\
&=\frac1\pi\Big[\frac{64}{81}(G-1)+\frac{4\pi^2}{243}+\frac{91\pi^4}{2160}\nonumber\\
&\qquad+\frac{35\,\zeta(3)}{18}+\frac{5\,\zeta(5)}{2}\Big],\nonumber\\
I_2&\equiv\int_{-\infty}^{\infty}d\nu'\,P(2,\nu')\,w\big(2,\tfrac12+i\nu'\big)\nonumber\\
&=\frac1\pi\Big[-\frac{32\,G}{81}-\frac{29}{648}+\frac{\zeta(3)}{72}+\frac{49\pi^2}{3888}\Big].
\label{eq:I02}
\end{align}
Combining the two equal integrals at nonzero conformal spin with $I_0$ cancels Catalan's constant,
\begin{align}
&\pi\,(I_0+2I_2)=\Big(\frac{64}{81}-2\cdot\frac{32}{81}\Big)G-\frac{64}{81}-2\cdot\frac{29}{648}\nonumber\\
&\quad+\Big(\frac{4}{243}+2\cdot\frac{49}{3888}\Big)\pi^2+\frac{91\pi^4}{2160}\nonumber\\
&\quad+\Big(\frac{35}{18}+2\cdot\frac{1}{72}\Big)\zeta(3)+\frac{5\,\zeta(5)}{2}\nonumber\\
&\quad=-\frac{95}{108}+\frac{\pi^2}{24}+\frac{91\pi^4}{2160}+\frac{71\,\zeta(3)}{36}+\frac{5\,\zeta(5)}{2}\nonumber\\
&\quad=\frac{91\pi^4+90\pi^2-1900+4260\,\zeta(3)+5400\,\zeta(5)}{2160}.
\label{eq:I0I2}
\end{align}
Division by the norm $Z_{1,0}$ of Eq.~\eqref{eq:Z} then gives Eq.~\eqref{eq:wP}. At nonzero $\nu$ the same residue calculation yields the numerator of $\langle w\rangle_P(\nu)$ along the $n=1$ eigenvalue. The individual singular terms cancel in the limit $\nu\to0$, leaving $I_0+2I_2$. The normalization tends to the value in Eq.~\eqref{eq:Zone}. Using the trigamma function $\psi'$, we obtain
\begin{widetext}
\begin{align}
Z_{1,\nu}&=-\frac{3\pi}{2}\,\frac{\chi(1,\nu)}{\nu^2(1+\nu^2)},\qquad Z_{1,0}=\lim_{\nu\to0}Z_{1,\nu}=3\pi\zeta(3),\nonumber\\
Z_{1,\nu}\,\langle w\rangle_P(\nu)&=\frac{\pi\big[26+115\nu^2+45\nu^4+4\nu^6-6\,(26+106\nu^2+35\nu^4+3\nu^6)\cosh(\pi\nu)/\sinh^2(\pi\nu)\big]}{6\,(4+\nu^2)(9+\nu^2)(1+4\nu^2)\,\nu^2(1+\nu^2)^2}\nonumber\\
&\quad-\frac{(-36+22\nu^2+11\nu^4+\nu^6)\big[\mathrm{Im}\,\psi'(\tfrac12+\tfrac{i\nu}{2})-\mathrm{Im}\,\psi'(1+\tfrac{i\nu}{2})\big]}{6\pi\,\nu^3(1+\nu^2)^2(4+\nu^2)(9+\nu^2)}\nonumber\\
&\quad+\frac{78+84\nu^2-8\nu^4-52\nu^6-15\nu^8-\nu^{10}}{3\pi\,\nu^4(1+\nu^2)^3(4+\nu^2)(9+\nu^2)}
+\frac{\zeta(3)}{\pi\,\nu^2(1+\nu^2)}.
\label{eq:Fnu}
\end{align}
\end{widetext}
The non-planar quark contribution at $n=1$ is $-(3/2)\langle w\rangle_P(\nu)$ in units of $n_f\alpha_s^2/N_c$. This function is even in $\nu$.
Its value at $\nu=0$, obtained by using the branch-point average of Eq.~\eqref{eq:wP}, gives the coefficient in Eq.~\eqref{eq:main}. In the same units the coefficient $a_q$ gives the non-planar quark contribution to the $\nu^2$ coefficient of Eq.~\eqref{eq:omegaexp}. Its defining expansion is
\begin{equation}
-\tfrac32\langle w\rangle_P(\nu)=-\tfrac32\langle w\rangle_P+a_q\,\nu^2+O(\nu^4).
\label{eq:Fexp}
\end{equation}

\section{Isolation of the BLV state}
\label{sm:levels}
We examine the LO three-gluon levels that enter the isolation condition for the first-order shift of the BLV eigenvalue in Sec.~\ref{sec:shift}. At the same conformal labels, every other level that the non-planar quark term can mix with the BLV state must have a different eigenvalue. The non-planar quark term acts with the same kernel and color coefficient on each pair, so permutations of the three gluons leave it invariant. Hence it mixes the BLV state only with levels of zero quasimomentum. For the LO spectrum we use the integral representation of Ref.~\cite{DerkachovKorchemskyKotanskiManashov:2002}. At fixed conformal weights this representation reduces the Baxter equation to a third-order differential equation. Requiring a single-valued combination of its solutions with those of the conjugate equation quantizes $q_3$. Each allowed value defines one level, with $\omega/\bar\alpha_s=-E_3/4$ for energy $E_3$ in that convention.

We seek the normalizable solutions with $q_3=0$ at the LO branch point $(n,\nu)=(1,0)$. We use the three-gluon wave functions and scalar product of Ref.~\cite{DerkachovKorchemskyKotanskiManashov:2002}. The wave function $\varphi$ depends on the complex transverse positions $x_1,x_2,x_3$ of the gluons. Each gluon carries holomorphic conformal weight 0 and antiholomorphic weight 1, while the state carries the weights $h=1$ and $\bar h=0$ at the reference point $x_0$ of Sec.~\ref{sec:formalism}. The squared norm in this scalar product is the integral of $|\varphi|^2$ over the three positions.

We write $x_{ij}=x_i-x_j$ for the gluon separations. The integrals of motion $q_3$ and $\bar q_3$ act on the holomorphic and antiholomorphic coordinates, respectively,
\begin{align}
q_3&=i^3\,x_{12}x_{23}x_{31}\,\partial_{x_1}\partial_{x_2}\partial_{x_3},\nonumber\\
\bar q_3\,\varphi&=i^3\,\partial_{\bar x_1}\partial_{\bar x_2}\partial_{\bar x_3}\big(\bar x_{12}\bar x_{23}\bar x_{31}\,\varphi\big).
\label{eq:q3charges}
\end{align}

The solutions with $q_3=0$ at these weights obey $q_3\varphi=0$ and $\bar q_3\varphi=0$. A solution of these two equations represents a state precisely when it is normalizable in this scalar product. The first equation determines the dependence of $\bar x_{12}\bar x_{23}\bar x_{31}\,\varphi$ on the gluon positions and the second determines its dependence on their complex conjugates. Writing $x_{i0}=x_i-x_0$ for the differences from the reference point, we obtain the respective bases
\begin{align}
&\Big\{r_{12},\ r_{23},\ \sum_{\rm cyc}r_{ij}\ln r_{ij}\Big\},\nonumber\\
&\Big\{1,\ \ln\frac{\bar r_{12}}{\bar r_{23}},\ \ln\frac{\bar r_{31}}{\bar r_{23}}\Big\},\qquad r_{ij}=\frac{x_{ij}}{x_{i0}x_{j0}}.
\label{eq:q3sol}
\end{align}
Here $\bar r_{ij}$ is the complex conjugate of $r_{ij}$. The three cyclic pairs obey $r_{12}+r_{23}+r_{31}=0$, so only two linear terms are independent. The general solution for $\bar x_{12}\bar x_{23}\bar x_{31}\,\varphi$ is a linear combination of the nine products of one function from each set.

Carrying one gluon around another changes the logarithms by multiples of $2\pi i$. Requiring the wave function to return to its initial value leaves exactly three independent combinations. The general combination and the limit that identifies its third term are
\begin{align}
\bar x_{12}\bar x_{23}\bar x_{31}\,\varphi&=c_1\,r_{12}+c_2\,r_{23}+c_3\sum_{\rm cyc}r_{ij}\ln|r_{ij}|^2,\nonumber\\
\sum_{\rm cyc}r_{ij}\ln|r_{ij}|^2&=\lim_{\nu\to0}\frac{1}{i\nu}\sum_{\rm cyc}E^{1,\nu}(x_i,x_j;x_0).
\label{eq:q3single}
\end{align}
The coefficients $c_1$, $c_2$ and $c_3$ are arbitrary. The wave function of the BLV state at $(1,\nu)$ is proportional to the cyclic sum of the conformal eigenfunctions of Eq.~\eqref{eq:Enu} divided by $\bar x_{12}\bar x_{23}\bar x_{31}$~\cite{BartelsLipatovVacca:2000,DerkachovKorchemskyKotanskiManashov:2002}. The sum vanishes at $\nu=0$, but dividing it by $i\nu$ before taking the limit gives the third function in the first line. The solutions with $q_3=0$ at $h=1$ were given in Ref.~\cite{Kotanski:2006} for another choice of the three-gluon wave functions.

We evaluate the norm in terms of the separation $x_{32}=x_3-x_2$ of gluons 2 and 3. For the position of gluon 1 we use the cross ratio $\kappa=x_{12}x_{30}/(x_{10}x_{32})$. Sending the reference point to infinity then gives $x_1=x_2+\kappa\,x_{32}$. The large-$|x_0|$ form of the integrand separates the squared norm into three integrals,
\begin{align}
&\int d^2x_1\,d^2x_2\,d^2x_3\,|\varphi|^2\nonumber\\
&\quad\to\frac{1}{|x_0|^4}\int d^2x_2\int\frac{d^2x_{32}}{|x_{32}|^2}\int\frac{d^2\kappa}{|1-\kappa|^2}\nonumber\\
&\qquad\times\Big|c_1-\frac{c_2}{\kappa}+c_3\Big[\ln|\kappa|^2+\frac{1-\kappa}{\kappa}\ln|1-\kappa|^2\Big]\Big|^2.
\label{eq:q3norm}
\end{align}
The integral over $x_2$ is the volume of translations, and the integral over the gluon separation is the volume of dilatations and rotations. These volumes are common to every solution at these labels. We set these common volumes aside and call a solution normalizable when its remaining integral over $\kappa$ converges.

The BLV state has $c_1=c_2=0$. Its integrand grows at most as the square of a logarithm near zero and one. At infinity it falls as $|\kappa|^{-4}$ up to logarithms, so the integral converges. These three limits correspond to coincidences of gluons 1 and 2, 1 and 3, and 2 and 3, respectively. A nonzero $c_2$ makes the integrand grow as $|\kappa|^{-2}$ near zero. The same power occurs at infinity when $c_2=0$ and $c_1\neq0$. Thus the norm of every single-valued solution other than a multiple of the BLV state diverges logarithmically where two gluons meet. We have shown that in the scalar product of Ref.~\cite{DerkachovKorchemskyKotanskiManashov:2002} only the BLV state has $q_3=0$ at $(n,\nu)=(1,0)$.

The mirror permutation reverses the gluon order. It takes a level at $q_3$ to its mirror level at $-q_3$ with the same eigenvalue and opposite quasimomentum. At $\nu=0$ complex conjugation of $q_3$ preserves both the eigenvalue and the quasimomentum. At $(n,\nu)=(1,0)$ every other level found in our search with $|q_3|<\nQthreeCutWide$ lay below the BLV eigenvalue of zero. The nearest level of zero quasimomentum lay $\nGapBose\,\bar\alpha_s$ below it. Table~\ref{tab:levels} lists the levels found with $|q_3|<\nQthreeCut$ and $\omega>-\nOmegaCut\,\bar\alpha_s$.
For $q_3=\nLevelQthreeA$ and $\nLevelQthreeC$ the energies agreed with those of Ref.~\cite{DerkachovKorchemskyKotanskiManashov:2002} to $\nLevelDiffPercentA$ and $\nLevelDiffPercentC$ percent, respectively. Equation~\eqref{eq:H0} makes the LO spectrum in units of $\bar\alpha_s$ independent of $N_c$. At the reference parameters the shift at $\nu=0$ was about $\nShiftGapPercent$ percent of the gap to the nearest level of zero quasimomentum.

\begin{table}[t]
\caption{LO three-gluon levels found in the search of Appendix~\ref{sm:levels} at $(n,\nu)=(1,0)$ that lie within $|q_3|<\nQthreeCut$ and $\omega>-\nOmegaCut\,\bar\alpha_s$. We use the integral of motion and quasimomentum of Ref.~\cite{DerkachovKorchemskyKotanskiManashov:2002}. The four rows below the BLV state comprise ten allowed nonzero values of $q_3$, obtained by solving the quantization conditions of Appendix~\ref{sm:methods} at $h=1$ first at $\nLevelDigits$ digits and again at $\nLevelDigitsHigh$. Each row groups levels related by reversing the gluon order or conjugating $q_3$, which have a common eigenvalue at $\nu=0$. Bose symmetry requires zero quasimomentum and, at nonzero $q_3$, the combination of a level with its mirror level at $-q_3$. The eigenvalue in the last column corresponds to the singularity at $j=1+\omega$.}
\label{tab:levels}
\centering
\begin{tabular}{lcc}
\hline
$q_3$ & Quasimomentum & $\omega/\bar\alpha_s$\\
\hline
$0$ & $0$ & $0$\\
$\pm\nLevelQthreeA$ & $\pm2\pi/3$ & $-\nLevelOmegaA$\\
$\pm(\nLevelQthreeReB\pm\nLevelQthreeImB\,i)$ & $\mp2\pi/3$ & $-\nLevelOmegaB$\\
$\pm\nLevelQthreeC$ & $0$ & $-\nLevelOmegaC$\\
$\pm\nLevelQthreeD\,i$ & $0$ & $-\nLevelOmegaD$\\
\hline
\end{tabular}
\end{table}
We followed the levels from the $|q_3|<\nQthreeCutWide$ search numerically as $\nu$ increased from zero along $n=1$. Among those found with zero quasimomentum, all gaps decreased over $0\leq\nu\leq\nNuEdge$ and remained positive. The closest levels came from $q_3=\nLevelQthreeC$ and its mirror level. Their gap decreased to $\nGapNuOne\,\bar\alpha_s$ at $\nu=1$ and to $\nGapNuThree\,\bar\alpha_s$ at $\nu=\nNuEdge$, the end of the lower panel of Fig.~\ref{fig:shift}. Just beyond that range these levels crossed the BLV eigenvalue at $\nu=\nNuCrossSame$. Levels of quasimomentum $\pm2\pi/3$ crossed the BLV eigenvalue within the range at $\nu=\nNuCrossOtherA$ and $\nu=\nNuCrossOtherB$. Their phase under cyclic permutation excludes mixing with the BLV state.

To compute levels of zero quasimomentum at $(n,\nu)=(1,0)$ beyond the search, we obtained starting values from the WKB formula of Ref.~\cite{DerkachovKorchemskyKotanskiManashov:2002} for the allowed values of $q_3$. We reached $|q_3|$ of about $\nQthreeLargeMax$. Each of these levels lay at least $\nOmegaLargeTop\,\bar\alpha_s$ below the BLV eigenvalue, as did all its partners under the symmetries that define a row of Table~\ref{tab:levels}.

We compared the energy $E_3$ of every level found at $(1,0)$, irrespective of quasimomentum, with $4\ln|q_3|+12\gamma_E$, where $\gamma_E$ is Euler's constant. We found numerically that the absolute difference was less than $\nDevMaxOne$ for $|q_3|>\nFitQMin$ and less than $\nDevMaxSix$ for $|q_3|>\nQthreeCutWide$. For the levels we found, the eigenvalue $\omega$ fell logarithmically as $|q_3|$ increased up to about $\nQthreeLargeMax$. We therefore expect levels at larger $|q_3|$ to lie further below the BLV eigenvalue.

\section{The planar NLO term and the scale dependence}
\label{sm:branch}
To compare the eigenvalues used in Sec.~\ref{sec:location} we need the change from the Wilson-line to the composite dipole operator. The composite operator differs by a gluon term of order $\alpha_s$ with the LO color structure, and the QCD evolution of the two operators has the same quark term~\cite{BalitskyChirilli:2009}. The resulting difference of NLO kernels is a commutator with the LO kernel. At zero momentum transfer the commutator contributes a term proportional to $\chi\chi'$ at fixed Mellin label. For $n=1$ this term is imaginary and odd on the line $\gamma=1/2+i\nu$, starting at order $\nu^3$. The $n=1$ eigenvalues of the composite and Wilson-line dipole operators thus have the same real part along the line and the same value at $\nu=0$ through NLO.

We denote the $\beta_0$ terms of the color-singlet NLO BFKL eigenvalue of Kotikov and Lipatov~\cite{KotikovLipatov:2000} by $\omega_\beta$ at the reference scale. At $n=1$ they take the form
\begin{equation}
\omega_\beta(\nu)=-\frac{\bar\alpha_s^2}{8}\,\frac{\beta_0}{N_c}\Big[\chi^2(1,\nu)-\psi'(1+i\nu)+\psi'(1-i\nu)\Big].
\label{eq:omegab}
\end{equation}
The difference of trigamma functions is odd and imaginary on this line. A square-root running-coupling factor in the eigenfunctions removes the difference~\cite{KotikovLipatov:2000}. The real term proportional to $\chi^2$ begins at order $\nu^4$. The LO curvature in Eq.~\eqref{eq:omegaexp} follows from Eq.~\eqref{eq:chi}. The planar NLO term involves the function $\Phi$ of Ref.~\cite{KotikovLipatov:2000}. At $n=1$ its defining integral converges for $\mathrm{Re}\,\gamma>-1/2$ and takes the form
\begin{align}
\Phi(1,\gamma)={}&-\int_0^1 d\xi\,\xi^{\gamma-3/2}\Big[\tfrac12\,\mathrm{Li}_2(\xi^2)\nonumber\\
&+\ln\xi\,\ln(1+\xi)\Big].
\label{eq:PhiKL}
\end{align}
This integral gives $\Phi''(1,1/2)=(11/2)\zeta(5)$, which we use in expanding Eq.~\eqref{eq:planar} for the planar part of the quadratic coefficient. The non-planar quark contribution to the $\nu^2$ coefficient follows from $a_q$ in Eq.~\eqref{eq:Fexp}.

The change of renormalization scale introduces a term $\omega_\mu$ that cancels the change of the LO eigenvalue through NLO. We include
\begin{equation}
\omega_\mu(\nu)=\bar\alpha_s^2(\mu^2)\,\frac{\beta_0}{4N_c}\,\ln\frac{\mu^2}{q^2}\,\chi(1,\nu)
\label{eq:omegamu}
\end{equation}
in the real eigenvalue, which reduces the magnitude of the LO curvature for $\mu^2<q^2$. At the reference parameters we find that the NLO terms cancel $\nCurvCancelPercent$ percent of the LO quadratic coefficient. In the fixed-scale evaluation of Sec.~\ref{sec:location} the total quadratic coefficient is negative down to about $\nMuThreshold\,q^2$ from $4q^2$ and positive at $q^2/4$. Evaluating the full eigenvalue at $\mu^2=q^2/4$ gives a maximum near $\nu=\nNuMaxLow$, about $\nExcessMaxLow$ above its value of $-\nShiftMuLowOrigin$ at zero.

\section{Numerical methods}
\label{sm:methods}
We give the numerical methods for the first-order shift and for the comparison of the BLV eigenvalue with the other three-gluon levels. We evaluated the integral for $\lambda_q$ in Eq.~\eqref{eq:Blambda} by Gauss-Legendre product quadrature in three regions, keeping the value from the finest grid for comparison with $4\pi^4 w$ at points where we repeated the quadrature with more nodes.
The integrand was evaluated at $\nKqDigits$ digits with nodes and weights in double precision.
In the search we found the nonzero $q_3$ levels directly at $h=1$ using the quantization conditions on the solutions of the Baxter equation. The integral representation of Ref.~\cite{DerkachovKorchemskyKotanskiManashov:2002} used for these conditions reduces the Baxter equation at fixed conformal weights to a third-order differential equation. The series about the singular point 0 gave the initial values for continuation around the singular point 1 by Taylor steps. We imposed single-valuedness of the bilinear combination of solutions of the equation and its conjugate~\cite{DerkachovKorchemskyKotanskiManashov:2002,JanikWosiek:1999} at $\nLevelDigits$ digits and repeated the evaluation at $\nLevelDigitsHigh$. Beyond the search the same method used starting values from the WKB formula of Ref.~\cite{DerkachovKorchemskyKotanskiManashov:2002} for the allowed $q_3$. At $q_3=0$ the conditions at $h=1$ are identities, so we used the limit of the closed form of the BLV eigenvalue there. For $\nu\ne0$ we followed the gaps of the levels found in the search by matching the series at the two singular points~\cite{DerkachovKorchemskyKotanskiManashov:2002} and joining the results to the direct values near zero. The steps of $\nFollowStep$ in $\nu$ were taken in double precision, with the sampled values confirmed at $\nFollowDigits$ digits. We computed the curves at $\nCurveDigits$ digits using the integral for $\Phi$ in Eq.~\eqref{eq:PhiKL}. The retained digits were stable under a change in working precision and, for the levels in the search, under a change of method.

The numerical comparisons below agree to the precision needed for each result.
The cubic color factors in Appendix~\ref{sm:color} were checked against traces of explicit SU(3), SU(4) and SU(5) generators. The coefficients in Eqs.~\eqref{eq:Bcross} and~\eqref{eq:Bbf} agree identically with the symbolic contractions for general $N_c$. The overlap in Eq.~\eqref{eq:single} that enters the distribution $P_{n,\nu}$ was compared with the value from direct integration on the cross-ratio plane at eight sets of real labels with odd $n+n'$ (conformal spin $n=1$ or $n=3$ for the three-gluon state and $n'=0,\pm2$ for the pair), including two sets at $(1,0)$. The overlap at even $n+n'$ vanished, with the overlaps at odd $n+n'$ setting the scale for the comparison with zero. The eigenvalue $\lambda_q$ from Eq.~\eqref{eq:Blambda} was compared with $4\pi^4 w$ at $n'=0,\pm2$ for the sampled $\nu'$ from $\nKqNuMin$ to $\nKqNuMax$. Its vanishing at $|n'|=4,6$ was tested relative to the values at $n'=0$.

The ratio of inner products was evaluated by direct integration over the $q\bar q$ positions and the cross ratio on three outer grids without the distribution $P_{1,0}$ or the closed forms of Sec.~\ref{sec:result} and Appendix~\ref{sm:closed}. The ratio in Eq.~\eqref{eq:expect} at $(n,\nu)=(1,0)$ and the corresponding ratios at $\nu=\nRatioNuA,\nRatioNuB$ were compared with $4\pi^4\langle w\rangle_P(\nu)$. The residue integrals in Eq.~\eqref{eq:I02} and the average in Eq.~\eqref{eq:wP} were checked by quadrature. The $\nu$ dependence in Eq.~\eqref{eq:Fnu} was tested by quadrature at $\nu=\nFnuNuA,\nFnuNuB,\nFnuNuC,\nFnuNuD$. The factor of two in Appendix~\ref{sm:color} and the coefficient in Eq.~\eqref{eq:main} were checked by evaluating the Abelian part of the symmetric adjoint pair kernel of Ref.~\cite{BartelsFadinLipatovVacca:2013} at zero momentum transfer. Its eigenvalue on the conformal partial wave $\Psi_{n',\nu'}$ was $-(n_f\alpha_s^2/2N_c)\,w(n',1/2+i\nu')$, twice the color-singlet value for every even $n'$ and real $\nu'$. The coefficient in Eq.~\eqref{eq:main} was compared with three times the average of this eigenvalue over $P_{1,0}$, with the eigenvalue expressed in units of $n_f\alpha_s^2/N_c$.

The contact coefficient in Eq.~\eqref{eq:S3contact}, positive at $y$ and negative at $x$, was checked by direct four-dimensional quadrature in the original coordinates, in which we paired $K_3$ symmetrically with each of three test functions and compared the integral with the coefficient times the nonzero difference in end values (value at $y$ minus value at $x$). The projectile integral in Eq.~\eqref{eq:transfer} was compared with the target integral at separated dipole ends. The two sides of Eq.~\eqref{eq:lefteig} for the non-dipole form $T(D_{xy},D_{zz'},D_{zz'})$ at $n=1$ agreed to $\nAgreeLeftEig$ significant digits at $\nu=\nLeftEigNu$.
At $\nu=0$ the left side vanished to this precision when measured against the overlap $\Pi_{1,0}$ of the form itself. The right side is zero there because $\chi(1,0)=0$.

Every nonzero $q_3$ level found in the search at $h=1$ was also obtained by continuing a fundamental system from the midpoint between 0 and 1 and imposing single-valuedness of its bilinear form. The values of $q_3$ and $E_3$ for each level beyond the search were compared at its two working precisions to check the range of $|q_3|$ and the lower bound on the gap to the BLV eigenvalue in Appendix~\ref{sm:levels}. The curves in Fig.~\ref{fig:shift} were checked by an independent evaluation at points spanning $0\leq\nu\leq1$ in the upper panel and at sampled values up to $\nu=\nNuEdge$ in the lower panel.

\bibliography{refs}

\begin{thebibliography}{19}%
\makeatletter
\providecommand \@ifxundefined [1]{%
 \@ifx{#1\undefined}
}%
\providecommand \@ifnum [1]{%
 \ifnum #1\expandafter \@firstoftwo
 \else \expandafter \@secondoftwo
 \fi
}%
\providecommand \@ifx [1]{%
 \ifx #1\expandafter \@firstoftwo
 \else \expandafter \@secondoftwo
 \fi
}%
\providecommand \natexlab [1]{#1}%
\providecommand \enquote  [1]{``#1''}%
\providecommand \bibnamefont  [1]{#1}%
\providecommand \bibfnamefont [1]{#1}%
\providecommand \citenamefont [1]{#1}%
\providecommand \href@noop [0]{\@secondoftwo}%
\providecommand \href [0]{\begingroup \@sanitize@url \@href}%
\providecommand \@href[1]{\@@startlink{#1}\@@href}%
\providecommand \@@href[1]{\endgroup#1\@@endlink}%
\providecommand \@sanitize@url [0]{\catcode `\\12\catcode `\$12\catcode
  `\&12\catcode `\#12\catcode `\^12\catcode `\_12\catcode `\%12\relax}%
\providecommand \@@startlink[1]{}%
\providecommand \@@endlink[0]{}%
\providecommand \url  [0]{\begingroup\@sanitize@url \@url }%
\providecommand \@url [1]{\endgroup\@href {#1}{\urlprefix }}%
\providecommand \urlprefix  [0]{URL }%
\providecommand \Eprint [0]{\href }%
\providecommand \doibase [0]{https://doi.org/}%
\providecommand \selectlanguage [0]{\@gobble}%
\providecommand \bibinfo  [0]{\@secondoftwo}%
\providecommand \bibfield  [0]{\@secondoftwo}%
\providecommand \translation [1]{[#1]}%
\providecommand \BibitemOpen [0]{}%
\providecommand \bibitemStop [0]{}%
\providecommand \bibitemNoStop [0]{.\EOS\space}%
\providecommand \EOS [0]{\spacefactor3000\relax}%
\providecommand \BibitemShut  [1]{\csname bibitem#1\endcsname}%
\let\auto@bib@innerbib\@empty
%</preamble>
\bibitem [{\citenamefont {Abazov}\ \emph {et~al.}(2021)\citenamefont {Abazov}
  \emph {et~al.}}]{D0TOTEM:2021}%
  \BibitemOpen
  \bibfield  {author} {\bibinfo {author} {\bibfnamefont {V.~M.}\ \bibnamefont
  {Abazov}} \emph {et~al.} (\bibinfo {collaboration} {D0, TOTEM}),\ }\bibfield
  {title} {\bibinfo {title} {{Odderon Exchange from Elastic Scattering
  Differences between $pp$ and $p \bar{p}$ Data at 1.96~TeV and from pp Forward
  Scattering Measurements}},\ }\href
  {https://doi.org/10.1103/PhysRevLett.127.062003} {\bibfield  {journal}
  {\bibinfo  {journal} {Phys. Rev. Lett.}\ }\textbf {\bibinfo {volume} {127}},\
  \bibinfo {pages} {062003} (\bibinfo {year} {2021})},\ \Eprint
  {https://arxiv.org/abs/2012.03981} {arXiv:2012.03981 [hep-ex]} \BibitemShut
  {NoStop}%
\bibitem [{\citenamefont {Kovchegov}\ \emph {et~al.}(2004)\citenamefont
  {Kovchegov}, \citenamefont {Szymanowski},\ and\ \citenamefont
  {Wallon}}]{KovchegovSzymanowskiWallon:2004}%
  \BibitemOpen
  \bibfield  {author} {\bibinfo {author} {\bibfnamefont {Y.~V.}\ \bibnamefont
  {Kovchegov}}, \bibinfo {author} {\bibfnamefont {L.}~\bibnamefont
  {Szymanowski}},\ and\ \bibinfo {author} {\bibfnamefont {S.}~\bibnamefont
  {Wallon}},\ }\bibfield  {title} {\bibinfo {title} {{Perturbative odderon in
  the dipole model}},\ }\href {https://doi.org/10.1016/j.physletb.2004.02.036}
  {\bibfield  {journal} {\bibinfo  {journal} {Phys. Lett. B}\ }\textbf
  {\bibinfo {volume} {586}},\ \bibinfo {pages} {267} (\bibinfo {year}
  {2004})},\ \Eprint {https://arxiv.org/abs/hep-ph/0309281}
  {arXiv:hep-ph/0309281} \BibitemShut {NoStop}%
\bibitem [{\citenamefont {Bartels}\ \emph {et~al.}(2000)\citenamefont
  {Bartels}, \citenamefont {Lipatov},\ and\ \citenamefont
  {Vacca}}]{BartelsLipatovVacca:2000}%
  \BibitemOpen
  \bibfield  {author} {\bibinfo {author} {\bibfnamefont {J.}~\bibnamefont
  {Bartels}}, \bibinfo {author} {\bibfnamefont {L.~N.}\ \bibnamefont
  {Lipatov}},\ and\ \bibinfo {author} {\bibfnamefont {G.~P.}\ \bibnamefont
  {Vacca}},\ }\bibfield  {title} {\bibinfo {title} {{A New odderon solution in
  perturbative QCD}},\ }\href {https://doi.org/10.1016/S0370-2693(00)00221-5}
  {\bibfield  {journal} {\bibinfo  {journal} {Phys. Lett. B}\ }\textbf
  {\bibinfo {volume} {477}},\ \bibinfo {pages} {178} (\bibinfo {year}
  {2000})},\ \Eprint {https://arxiv.org/abs/hep-ph/9912423}
  {arXiv:hep-ph/9912423} \BibitemShut {NoStop}%
\bibitem [{\citenamefont {Kovchegov}\ and\ \citenamefont
  {Sievert}(2012)}]{KovchegovSievert:2012}%
  \BibitemOpen
  \bibfield  {author} {\bibinfo {author} {\bibfnamefont {Y.~V.}\ \bibnamefont
  {Kovchegov}}\ and\ \bibinfo {author} {\bibfnamefont {M.~D.}\ \bibnamefont
  {Sievert}},\ }\bibfield  {title} {\bibinfo {title} {{New mechanism for
  generating a single transverse spin asymmetry}},\ }\href
  {https://doi.org/10.1103/PhysRevD.86.034028} {\bibfield  {journal} {\bibinfo
  {journal} {Phys. Rev. D}\ }\textbf {\bibinfo {volume} {86}},\ \bibinfo
  {pages} {034028} (\bibinfo {year} {2012})},\ \bibinfo {note} {[Erratum: Phys.
  Rev. D {\bf 86}, 079906 (2012)]},\ \Eprint {https://arxiv.org/abs/1201.5890}
  {arXiv:1201.5890 [hep-ph]} \BibitemShut {NoStop}%
\bibitem [{\citenamefont {Caron-Huot}(2015)}]{CaronHuot:2015}%
  \BibitemOpen
  \bibfield  {author} {\bibinfo {author} {\bibfnamefont {S.}~\bibnamefont
  {Caron-Huot}},\ }\bibfield  {title} {\bibinfo {title} {{When does the gluon
  reggeize?}},\ }\href {https://doi.org/10.1007/JHEP05(2015)093} {\bibfield
  {journal} {\bibinfo  {journal} {JHEP}\ }\textbf {\bibinfo {volume}
  {2015}}\bibfield  {number} {\bibinfo  {number} { (05)},\ \bibinfo {pages}
  {093}},\ }\Eprint {https://arxiv.org/abs/1309.6521} {arXiv:1309.6521
  [hep-th]} \BibitemShut {NoStop}%
\bibitem [{\citenamefont {Sta{\'s}to}(2009)}]{Stasto:2009}%
  \BibitemOpen
  \bibfield  {author} {\bibinfo {author} {\bibfnamefont {A.~M.}\ \bibnamefont
  {Sta{\'s}to}},\ }\bibfield  {title} {\bibinfo {title} {{Small x resummation
  and the Odderon}},\ }\href {https://doi.org/10.1016/j.physletb.2009.07.053}
  {\bibfield  {journal} {\bibinfo  {journal} {Phys. Lett. B}\ }\textbf
  {\bibinfo {volume} {679}},\ \bibinfo {pages} {288} (\bibinfo {year}
  {2009})},\ \Eprint {https://arxiv.org/abs/0904.4124} {arXiv:0904.4124
  [hep-ph]} \BibitemShut {NoStop}%
\bibitem [{\citenamefont {Kovchegov}(2013)}]{Kovchegov:2013}%
  \BibitemOpen
  \bibfield  {author} {\bibinfo {author} {\bibfnamefont {Y.~V.}\ \bibnamefont
  {Kovchegov}},\ }\bibfield  {title} {\bibinfo {title} {{Running Coupling
  Evolution for Diffractive Dissociation and the NLO Odderon Intercept}},\
  }\href {https://doi.org/10.1063/1.4802180} {\bibfield  {journal} {\bibinfo
  {journal} {AIP Conf. Proc.}\ }\textbf {\bibinfo {volume} {1523}},\ \bibinfo
  {pages} {335} (\bibinfo {year} {2013})},\ \Eprint
  {https://arxiv.org/abs/1212.2113} {arXiv:1212.2113 [hep-ph]} \BibitemShut
  {NoStop}%
\bibitem [{\citenamefont {Bartels}\ and\ \citenamefont
  {Vacca}(2013)}]{BartelsVacca:2013}%
  \BibitemOpen
  \bibfield  {author} {\bibinfo {author} {\bibfnamefont {J.}~\bibnamefont
  {Bartels}}\ and\ \bibinfo {author} {\bibfnamefont {G.~P.}\ \bibnamefont
  {Vacca}},\ }\bibfield  {title} {\bibinfo {title} {{Generalized Bootstrap
  Equations and possible implications for the NLO Odderon}},\ }\href
  {https://doi.org/10.1140/epjc/s10052-013-2602-8} {\bibfield  {journal}
  {\bibinfo  {journal} {Eur. Phys. J. C}\ }\textbf {\bibinfo {volume} {73}},\
  \bibinfo {pages} {2602} (\bibinfo {year} {2013})},\ \Eprint
  {https://arxiv.org/abs/1307.3985} {arXiv:1307.3985 [hep-th]} \BibitemShut
  {NoStop}%
\bibitem [{\citenamefont {Balitsky}\ and\ \citenamefont
  {Grabovsky}(2015)}]{BalitskyGrabovsky:2015}%
  \BibitemOpen
  \bibfield  {author} {\bibinfo {author} {\bibfnamefont {I.}~\bibnamefont
  {Balitsky}}\ and\ \bibinfo {author} {\bibfnamefont {A.~V.}\ \bibnamefont
  {Grabovsky}},\ }\bibfield  {title} {\bibinfo {title} {{NLO evolution of
  3-quark Wilson loop operator}},\ }\href
  {https://doi.org/10.1007/JHEP01(2015)009} {\bibfield  {journal} {\bibinfo
  {journal} {JHEP}\ }\textbf {\bibinfo {volume} {2015}}\bibfield  {number}
  {\bibinfo  {number} { (01)},\ \bibinfo {pages} {009}},\ }\Eprint
  {https://arxiv.org/abs/1405.0443} {arXiv:1405.0443 [hep-ph]} \BibitemShut
  {NoStop}%
\bibitem [{\citenamefont {Balitsky}\ and\ \citenamefont
  {Chirilli}(2008)}]{BalitskyChirilli:2008}%
  \BibitemOpen
  \bibfield  {author} {\bibinfo {author} {\bibfnamefont {I.}~\bibnamefont
  {Balitsky}}\ and\ \bibinfo {author} {\bibfnamefont {G.~A.}\ \bibnamefont
  {Chirilli}},\ }\bibfield  {title} {\bibinfo {title} {{Next-to-leading order
  evolution of color dipoles}},\ }\href
  {https://doi.org/10.1103/PhysRevD.77.014019} {\bibfield  {journal} {\bibinfo
  {journal} {Phys. Rev. D}\ }\textbf {\bibinfo {volume} {77}},\ \bibinfo
  {pages} {014019} (\bibinfo {year} {2008})},\ \Eprint
  {https://arxiv.org/abs/0710.4330} {arXiv:0710.4330 [hep-ph]} \BibitemShut
  {NoStop}%
\bibitem [{\citenamefont {Hatta}\ \emph {et~al.}(2005)\citenamefont {Hatta},
  \citenamefont {Iancu}, \citenamefont {Itakura},\ and\ \citenamefont
  {McLerran}}]{HattaIancuItakuraMcLerran:2005}%
  \BibitemOpen
  \bibfield  {author} {\bibinfo {author} {\bibfnamefont {Y.}~\bibnamefont
  {Hatta}}, \bibinfo {author} {\bibfnamefont {E.}~\bibnamefont {Iancu}},
  \bibinfo {author} {\bibfnamefont {K.}~\bibnamefont {Itakura}},\ and\ \bibinfo
  {author} {\bibfnamefont {L.}~\bibnamefont {McLerran}},\ }\bibfield  {title}
  {\bibinfo {title} {{Odderon in the color glass condensate}},\ }\href
  {https://doi.org/10.1016/j.nuclphysa.2005.05.163} {\bibfield  {journal}
  {\bibinfo  {journal} {Nucl. Phys. A}\ }\textbf {\bibinfo {volume} {760}},\
  \bibinfo {pages} {172} (\bibinfo {year} {2005})},\ \Eprint
  {https://arxiv.org/abs/hep-ph/0501171} {arXiv:hep-ph/0501171} \BibitemShut
  {NoStop}%
\bibitem [{\citenamefont {Kotikov}\ and\ \citenamefont
  {Lipatov}(2000)}]{KotikovLipatov:2000}%
  \BibitemOpen
  \bibfield  {author} {\bibinfo {author} {\bibfnamefont {A.~V.}\ \bibnamefont
  {Kotikov}}\ and\ \bibinfo {author} {\bibfnamefont {L.~N.}\ \bibnamefont
  {Lipatov}},\ }\bibfield  {title} {\bibinfo {title} {{NLO corrections to the
  BFKL equation in QCD and in supersymmetric gauge theories}},\ }\href
  {https://doi.org/10.1016/S0550-3213(00)00329-1} {\bibfield  {journal}
  {\bibinfo  {journal} {Nucl. Phys. B}\ }\textbf {\bibinfo {volume} {582}},\
  \bibinfo {pages} {19} (\bibinfo {year} {2000})},\ \Eprint
  {https://arxiv.org/abs/hep-ph/0004008} {arXiv:hep-ph/0004008} \BibitemShut
  {NoStop}%
\bibitem [{\citenamefont {Derkachov}\ \emph {et~al.}(2002)\citenamefont
  {Derkachov}, \citenamefont {Korchemsky}, \citenamefont {Kota{\'n}ski},\ and\
  \citenamefont {Manashov}}]{DerkachovKorchemskyKotanskiManashov:2002}%
  \BibitemOpen
  \bibfield  {author} {\bibinfo {author} {\bibfnamefont {S.~E.}\ \bibnamefont
  {Derkachov}}, \bibinfo {author} {\bibfnamefont {G.~P.}\ \bibnamefont
  {Korchemsky}}, \bibinfo {author} {\bibfnamefont {J.}~\bibnamefont
  {Kota{\'n}ski}},\ and\ \bibinfo {author} {\bibfnamefont {A.~N.}\ \bibnamefont
  {Manashov}},\ }\bibfield  {title} {\bibinfo {title} {{Noncompact Heisenberg
  spin magnets from high-energy QCD. II. Quantization conditions and energy
  spectrum}},\ }\href {https://doi.org/10.1016/S0550-3213(02)00842-8}
  {\bibfield  {journal} {\bibinfo  {journal} {Nucl. Phys. B}\ }\textbf
  {\bibinfo {volume} {645}},\ \bibinfo {pages} {237} (\bibinfo {year}
  {2002})},\ \Eprint {https://arxiv.org/abs/hep-th/0204124}
  {arXiv:hep-th/0204124} \BibitemShut {NoStop}%
\bibitem [{\citenamefont {Bartels}\ \emph {et~al.}(2013)\citenamefont
  {Bartels}, \citenamefont {Fadin}, \citenamefont {Lipatov},\ and\
  \citenamefont {Vacca}}]{BartelsFadinLipatovVacca:2013}%
  \BibitemOpen
  \bibfield  {author} {\bibinfo {author} {\bibfnamefont {J.}~\bibnamefont
  {Bartels}}, \bibinfo {author} {\bibfnamefont {V.~S.}\ \bibnamefont {Fadin}},
  \bibinfo {author} {\bibfnamefont {L.~N.}\ \bibnamefont {Lipatov}},\ and\
  \bibinfo {author} {\bibfnamefont {G.~P.}\ \bibnamefont {Vacca}},\ }\bibfield
  {title} {\bibinfo {title} {{NLO Corrections to the kernel of the
  BKP-equations}},\ }\href {https://doi.org/10.1016/j.nuclphysb.2012.10.024}
  {\bibfield  {journal} {\bibinfo  {journal} {Nucl. Phys. B}\ }\textbf
  {\bibinfo {volume} {867}},\ \bibinfo {pages} {827} (\bibinfo {year}
  {2013})},\ \Eprint {https://arxiv.org/abs/1210.0797} {arXiv:1210.0797
  [hep-ph]} \BibitemShut {NoStop}%
\bibitem [{\citenamefont {Balitsky}\ and\ \citenamefont
  {Chirilli}(2009)}]{BalitskyChirilli:2009}%
  \BibitemOpen
  \bibfield  {author} {\bibinfo {author} {\bibfnamefont {I.}~\bibnamefont
  {Balitsky}}\ and\ \bibinfo {author} {\bibfnamefont {G.~A.}\ \bibnamefont
  {Chirilli}},\ }\bibfield  {title} {\bibinfo {title} {{NLO evolution of color
  dipoles in $\mathcal{N}=4$ SYM}},\ }\href
  {https://doi.org/10.1016/j.nuclphysb.2009.07.003} {\bibfield  {journal}
  {\bibinfo  {journal} {Nucl. Phys. B}\ }\textbf {\bibinfo {volume} {822}},\
  \bibinfo {pages} {45} (\bibinfo {year} {2009})},\ \Eprint
  {https://arxiv.org/abs/0903.5326} {arXiv:0903.5326 [hep-ph]} \BibitemShut
  {NoStop}%
\bibitem [{\citenamefont {Kovchegov}\ and\ \citenamefont
  {Santiago}(2022)}]{KovchegovSantiago:2022}%
  \BibitemOpen
  \bibfield  {author} {\bibinfo {author} {\bibfnamefont {Y.~V.}\ \bibnamefont
  {Kovchegov}}\ and\ \bibinfo {author} {\bibfnamefont {M.~G.}\ \bibnamefont
  {Santiago}},\ }\bibfield  {title} {\bibinfo {title} {{T-odd leading-twist
  quark TMDs at small x}},\ }\href {https://doi.org/10.1007/JHEP11(2022)098}
  {\bibfield  {journal} {\bibinfo  {journal} {JHEP}\ }\textbf {\bibinfo
  {volume} {2022}}\bibfield  {number} {\bibinfo  {number} { (11)},\ \bibinfo
  {pages} {098}},\ }\Eprint {https://arxiv.org/abs/2209.03538}
  {arXiv:2209.03538 [hep-ph]} \BibitemShut {NoStop}%
\bibitem [{\citenamefont {Bartels}\ \emph {et~al.}(2020)\citenamefont
  {Bartels}, \citenamefont {Contreras},\ and\ \citenamefont
  {Vacca}}]{BartelsContrerasVacca:2020}%
  \BibitemOpen
  \bibfield  {author} {\bibinfo {author} {\bibfnamefont {J.}~\bibnamefont
  {Bartels}}, \bibinfo {author} {\bibfnamefont {C.}~\bibnamefont {Contreras}},\
  and\ \bibinfo {author} {\bibfnamefont {G.~P.}\ \bibnamefont {Vacca}},\
  }\bibfield  {title} {\bibinfo {title} {{The Odderon in QCD with running
  coupling}},\ }\href {https://doi.org/10.1007/JHEP04(2020)183} {\bibfield
  {journal} {\bibinfo  {journal} {JHEP}\ }\textbf {\bibinfo {volume}
  {2020}}\bibfield  {number} {\bibinfo  {number} { (04)},\ \bibinfo {pages}
  {183}},\ }\Eprint {https://arxiv.org/abs/1910.04588} {arXiv:1910.04588
  [hep-th]} \BibitemShut {NoStop}%
\bibitem [{\citenamefont {Janik}\ and\ \citenamefont
  {Wosiek}(1999)}]{JanikWosiek:1999}%
  \BibitemOpen
  \bibfield  {author} {\bibinfo {author} {\bibfnamefont {R.~A.}\ \bibnamefont
  {Janik}}\ and\ \bibinfo {author} {\bibfnamefont {J.}~\bibnamefont {Wosiek}},\
  }\bibfield  {title} {\bibinfo {title} {{A Solution of the Odderon Problem}},\
  }\href {https://doi.org/10.1103/PhysRevLett.82.1092} {\bibfield  {journal}
  {\bibinfo  {journal} {Phys. Rev. Lett.}\ }\textbf {\bibinfo {volume} {82}},\
  \bibinfo {pages} {1092} (\bibinfo {year} {1999})},\ \Eprint
  {https://arxiv.org/abs/hep-th/9802100} {arXiv:hep-th/9802100} \BibitemShut
  {NoStop}%
\bibitem [{\citenamefont {Kota{\'n}ski}(2006)}]{Kotanski:2006}%
  \BibitemOpen
  \bibfield  {author} {\bibinfo {author} {\bibfnamefont {J.}~\bibnamefont
  {Kota{\'n}ski}},\ }\bibfield  {title} {\bibinfo {title} {{Three particle
  Pomeron and odderon states in QCD}},\ }\href@noop {} {\bibfield  {journal}
  {\bibinfo  {journal} {Acta Phys. Polon. B}\ }\textbf {\bibinfo {volume}
  {37}},\ \bibinfo {pages} {2615} (\bibinfo {year} {2006})},\ \Eprint
  {https://arxiv.org/abs/hep-th/0603238} {arXiv:hep-th/0603238} \BibitemShut
  {NoStop}%
\end{thebibliography}%

\end{document}